\documentclass[twocolumn, twocolappendix]{aastex631}
\usepackage{amsmath}
\let\originaltablenum\tablenum
\let\tablenum\undefined
\usepackage{siunitx}
\let\tablenum\originaltablenum
\usepackage{hyperref}
\usepackage{textcomp}
\usepackage[table]{xcolor}

\let\olditemize\itemize
\renewcommand{\itemize}{\olditemize\setlength{\itemsep}{1pt}\setlength{\parsep}{0pt}\setlength{\parskip}{0pt}}

\begin{document}

\title{The thermal and kinematic Sunyaev-Zeldovich effect in galaxy clusters and filaments using multifrequency temperature maps of the cosmic microwave background: Abell 399--Abell 401 cluster pair case study}

\author[0000-0002-3937-4662]{Ajay~S.~Gill} \affiliation{Department of Aeronautics \& Astronautics, Massachusetts Institute of Technology, 77 Mass. Avenue, Cambridge, MA 02139, USA}

\author[0000-0002-1697-3080]{Yilun~Guan}
\affiliation{Dunlap Institute for Astronomy and Astrophysics, University of Toronto, 50 St. George St., Toronto, ON M5S 3H4, Canada}

\author[0000-0003-1690-6678]{Adam~D.~Hincks} 
\affiliation{David A. Dunlap Dept of Astronomy and Astrophysics, University of Toronto, 50 St George Street, Toronto ON, M5S 3H4, Canada} 
\affiliation{Specola Vaticana (Vatican Observatory), V-00120 Vatican City State}

\author[0000-0003-3816-5372]{Tony~Mroczkowski} 
\affiliation{Institute of Space Sciences (CSIC-ICE), Carrer de Can Magrans, s/n, 08193 Cerdanyola del Vall\`es, Barcelona, Spain}
\affiliation{European Southern Observatory (ESO), Karl-Schwarzschild-Str. 2, D-85748, Garching, Germany}
 
\author[0000-0002-2287-1603]{Zachary~Atkins}
\affiliation{Department of Physics and Astronomy, University of Pennsylvania, 209 South 33rd Street, Philadelphia, PA 19104, USA}
\affiliation{Joseph Henry Laboratories of Physics, Jadwin Hall, Princeton University, Princeton, NJ, USA 08544}

\author[0000-0002-2011-9012]{Eleonora~Barbavara}
\affiliation{Dipartimento di Fisica, Sapienza Università di Roma, Piazzale Aldo Moro 5, I-00185 Roma, Italy}
\affiliation{Dipartimento di Fisica, Università di Roma Tor Vergata, Via della Ricerca Scientifica 1, I-00133, Roma, Italy}
\affiliation{David A. Dunlap Dept of Astronomy and Astrophysics, University of Toronto, 50 St George Street, Toronto ON, M5S 3H4, Canada}

\author[0000-0001-5210-7625]{Elia~S.~Battistelli}  
\affiliation{Dipartimento di Fisica, Sapienza Università di Roma, Piazzale Aldo
Moro 5, I-00185 Roma, Italy}

\author[0000-0003-2358-9949]{J.~Richard~Bond}
\affiliation{Canadian Institute for Theoretical Astrophysics, 60 St. George Street, 
University of Toronto, Toronto, ON, M5S 3H8, Canada}

\author[0000-0002-1297-3673]{William~Coulton}  
\affiliation{Kavli Institute for Cosmology Cambridge, Madingley Road, Cambridge CB3 0HA, UK}
\affiliation{DAMTP, Centre for Mathematical Sciences, University of Cambridge, Wilberforce Road, Cambridge CB3 OWA, UK}

\author[0000-0003-2856-2382]{Adri~J~Duivenvoorden} 
\affiliation{Max-Planck-Institut fur Astrophysik, Karl-Schwarzschild-Str. 1, 85748 Garching, Germany}

\author[0000-0002-8490-8117]{Matt~Hilton} \affiliation{Wits Centre for Astrophysics, School of Physics, University of the Witwatersrand, Private Bag 3, 2050, Johannesburg, South Africa} \affiliation{Astrophysics Research Centre, School of Mathematics, Statistics and Computer Science, University of KwaZulu-Natal, Durban 4001, South Africa}

\author[0000-0002-8816-6800]{John~P.~Hughes} 
\affiliation{Department of Physics and Astronomy, Rutgers, The State University of New Jersey, Piscataway, NJ USA 08854-8019}

\author[0000-0002-8458-0588]{Giovanni~Isopi}  
\affiliation{Dipartimento di Fisica, Sapienza Università di Roma, Piazzale Aldo
Moro 5, I-00185 Roma, Italy}

\author[0000-0001-9830-3103]{Joshiwa~van~Marrewijk} 
\affiliation{Leiden Observatory, Leiden University, P.O. Box 9513, 2300 RA Leiden, The Netherlands}

\author[0000-0001-6606-7142]{Kavilan~Moodley} 
\affiliation{Astrophysics Research Centre, University of KwaZulu-Natal, Westville Campus, Durban 4041, South Africa}
\affiliation{School of Mathematics, Statistics \& Computer Science, University of KwaZulu-Natal, Westville Campus, Durban 4041, South Africa}

\author[0000-0002-4478-7111]{Sigurd~Naess}  
\affiliation{Institute of Theoretical Astrophysics, University of Oslo, Norway}

\author[0000-0001-6541-9265]{Bruce~Partridge} 
\affiliation{Department of Physics and Astronomy, Haverford College, Haverford, PA, USA 19041}

\author[0000-0002-0418-6258]{Bernardita~Ried~Guachalla} 
\affiliation{Department of Physics, Stanford University, 382 Via Pueblo Mall, Stanford, CA 94305, USA}
\affiliation{Kavli Institute for Particle Astrophysics \& Cosmology, 452 Lomita Mall, Stanford, CA 94305, USA}
\affiliation{SLAC National Accelerator Laboratory, 2575 Sand Hill Road, Menlo Park, CA 94025, USA}

\author[0000-0003-1842-8104]{John~Orlowski-Scherer} 
\affiliation{Department of Physics and Astronomy, University of Pennsylvania, 209 South 33rd Street, Philadelphia, PA, USA 19104}

\author[0000-0002-8149-1352]{Crist\'obal Sif\'on}
\affiliation{Instituto de F\'isica, Pontificia Universidad Cat\'olica de Valpara\'iso, Casilla 4059, Valpara\'iso, Chile}

\author[0000-0002-2105-7589]{Eve~M.~Vavagiakis } 
\affiliation{Department of Physics, Duke University, Durham, NC, 27704}
\affiliation{Department of Physics, Cornell University, Ithaca, NY 14850}

\author[0000-0002-7567-4451]{Edward~J.~Wollack } 
\affiliation{NASA Goddard Space Flight Center, 8800 Greenbelt Road, Greenbelt, MD 20771, USA}

\correspondingauthor{Ajay~S.~Gill}
\email{agill.astro@gmail.com}

\begin{abstract}
We present a multifrequency and multi-instrument methodology to study the physical properties of galaxy clusters and cosmic filaments using cosmic microwave background observations. Our approach enables simultaneous measurement of both the thermal (tSZ) and kinematic Sunyaev-Zeldovich (kSZ) effects, incorporates relativistic corrections, and models astrophysical foregrounds such as thermal dust emission. We do this by jointly fitting a single physical model across multiple maps from multiple instruments at different frequencies, rather than fitting a model to a single Compton-$y$ map. We demonstrate the success of this method by fitting the Abell 399--Abell 401 galaxy cluster pair and filament system using archival data from the \textit{Planck} satellite and new, targeted deep data from the Atacama Cosmology Telescope, covering 11 different frequencies over 14 maps from 30\,GHz to 545\,GHz. Our tSZ results are consistent with previous work using Compton-\textit{y} maps. We measure the line-of-sight peculiar velocities of the cluster--filament system using the kSZ effect and find statistical uncertainties on individual cluster peculiar velocities of  $\lesssim\,$600\,km\,s$^{-1}$, which are competitive with current state-of-the-art measurements. Additionally, we measure the optical depth of the filament component with a signal-to-noise of 8.5$\sigma$ and reveal hints of its morphology. This modular approach is well-suited for application to future instruments across a wide range of millimeter and sub-millimeter wavebands.
\end{abstract}

\section{Introduction}
The cosmic microwave background (CMB) is the radiation from the early Universe released around 380,000 years after the Big Bang. First discovered in the mid-1960s \citep{cmb_discovery}, its analysis has played a pivotal role in enhancing our understanding of what the Universe is made of and how it has evolved over time. The CMB has proven to be a powerful tool for probing the large-scale structure of the Universe. One way this is achieved is by studying the interaction of CMB photons with hot electrons as they pass through galaxy clusters, the largest gravitationally bound objects in the Universe, with masses of order 10$^{13}$--10$^{15}$ $M_{\odot}$ and average intergalactic gas temperatures as high as 5--10\,keV. 

As CMB photons interact with the hot, ionized electrons in the intracluster medium, they gain energy via inverse-Compton scattering. This process creates a distortion in the CMB blackbody frequency spectrum towards higher frequencies, an effect known as the thermal Sunyaev-Zeldovich effect (tSZ; \citealt{z_s_1969, sz1970, sz1972}). The upscattering of photons by the tSZ effect results in a temperature decrement (increment) relative to the CMB blackbody at frequencies below (above) 217\,GHz.

In addition to the tSZ effect, the bulk motion of galaxy clusters also imparts a Doppler boost to the CMB photons. This effect, referred to as the kinematic Sunyaev-Zeldovich effect (kSZ), does not distort the shape of the blackbody spectrum but alters its apparent temperature as the photons thermalize in the rest frame of the moving cluster. The magnitude of the kSZ effect depends linearly on the line-of-sight optical depth through the galaxy cluster and its peculiar velocity relative to the Hubble flow. On the other hand, the tSZ effect depends linearly on the integrated pressure along the line of sight. \citealt{1999Birkinshaw} and \citealt{2019Mroczkowski} provide reviews of the SZ effects, with the latter focusing on more recent astrophysical applications as well as introducing some of the more recent relativistic corrections discussed in Section \ref{sec:methods}.

 The kSZ effect allows for measuring the peculiar velocities of galaxy clusters along the line of sight, probing the velocity field of the large-scale structure in the Universe, which is sensitive to cosmological parameters. In particular, the peculiar velocity distribution measured from samples of galaxy clusters can be used to measure the dark energy equation-of-state parameter \citep{2008Bhattacharya} and distinguish between models of dark energy and modified gravity \citep{2009Kosowsky}. 

\begin{figure}
    \centering
    \includegraphics[width=\columnwidth]{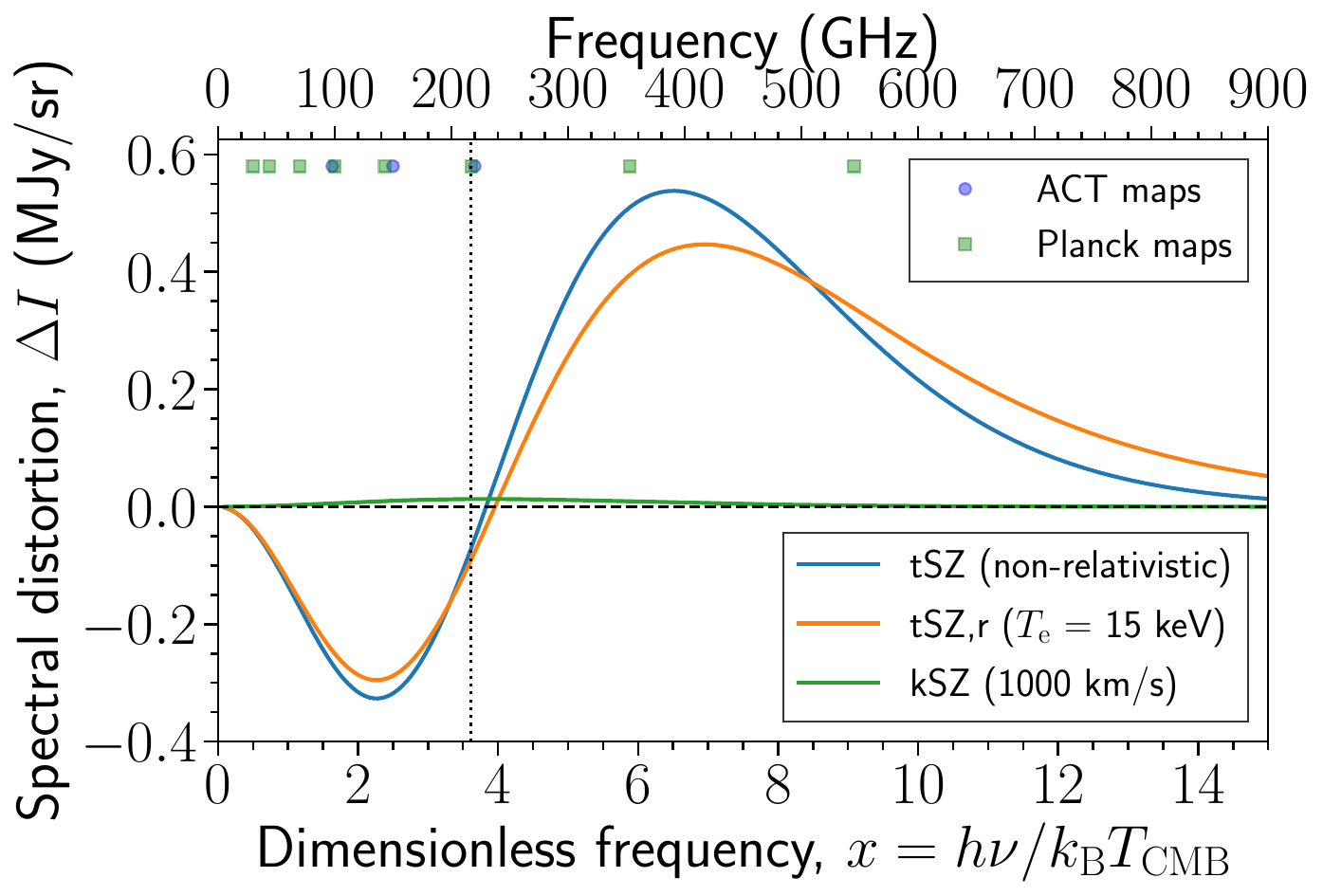}
    \caption{Thermal and kinematic SZ effect including relativistic corrections. We assume an optical depth of $\tau_{\rm e} = 0.01$ for these calculations. The vertical dotted line is at the tSZ null at 217\,GHz. For illustration, the vertical lines show the bands of the ACT and \textit{Planck} frequency maps we used in this work.}
    \label{fig:sz_compare}
\end{figure}

Despite the promise of kSZ surveys, measurements of the kSZ signal have proven to be an observational challenge. The kSZ signal has the same spectral dependence as the primary CMB anisotropies, and it is also weak: for instance, a galaxy cluster with an optical depth $\tau_{\rm e} = 0.005$ and a line-of-sight, receding peculiar velocity of 300 km s$^{-1}$ has a kSZ signal of $\Delta T_{\rm kSZ} \sim -13.6\, \mu$K. On the other hand, the primary CMB anisotropy on angular scales of $\sim 1^{\circ}$ is $\sim 70$ $\mu$K, about five times larger than the kSZ signal. Another challenge is adequately separating the tSZ effect from the kSZ effect. To do so, observations near tSZ null (217\,GHz) are important, where only the kSZ signal is present. Figure~\ref{fig:sz_compare} compares the tSZ and kSZ spectra for a fixed optical depth of 0.01. Relativistic effects, which become important when the gas is hot, alter the SZ spectra---particularly the tSZ spectrum (see Figure~\ref{fig:sz_compare})---and shift the tSZ null frequency. They must be accounted for to accurately model the SZ signal. Furthermore, the thermal emission of dust in member galaxies of the cluster must be accurately accounted for so as not to contaminate the kSZ and tSZ signals.

While the kSZ effect has now been detected with high significance by stacking or combining multiple galaxy clusters \citep{2021Schaan, 2024Hadzhiyska, 2025Bernardita}, attempts to constrain the kSZ effect in individual clusters remain limited. \citet{2013Sayers} measured the velocity of one of the sub-clusters in MACS J0717.5+3745, a triple merging cluster, to be $+2550 \pm 1050$ km s$^{-1}$ using 140 and 268\,GHz Bolocam data. \cite{2017_Adam} also measured the gas velocity within MACS J0717.5+3745 using the 150 and 260\,GHz bands of the NIKA camera at the IRAM 30-m telescope. \citet{sayers2019} measured the tSZ and kSZ signal in a sample of 10 merging galaxy clusters using data from Bolocam/AzTEC, \textit{Herschel}-SPIRE, \textit{Planck}, \textit{Chandra}, and the \textit{Hubble Space Telescope}, finding a median velocity uncertainty per cluster of $\sim 700$ km s$^{-1}$, but without detecting any non-zero peculiar velocities.

In addition to probing galaxy clusters, the SZ effect can also be used to study the material present between the clusters. The large-scale structure takes the form of a cosmic web, with galaxy clusters at the nodes connected by cosmic filaments. Simulations suggest that the diffuse gas in cosmic filaments contains around 30\% of the baryons in the Universe (e.g., \citealt{bond1996, 1999Cen, Dave_2001,  2021Espinosa, 2021Tuominen}). This cosmic web structure has been observationally verified by galaxy surveys (e.g., \citealt{1989Geller, colless2001, 2005Gott, 2008Sousbie, 2014Tempel}), molecular absorption lines (e.g., \citealt{2016Tejos, 2018Nicastro, 2018Pessa, 2019Nevalainen, 2021Bouma}), dispersion measurements of fast radio bursts (e.g., \citealt{2020Macquart}), and X-ray emission (e.g., \citealt{2015Eckert, 2020Tanimura, 2022Tanimura, Dietl2024}). Most relevant to our work, filaments have also been detected via stacking tSZ measurements from multiple clusters (e.g., \citealt{2019Graaff, 2019Tanimura, 2024Isopi, Lokken:2025}), via stacking of kSZ measurements \citep{Hadzhiyska:2025}, and in a few limited cases, detected or constrained in tSZ measurements of individual cluster pairs \citep{2013PlanckFilaments, 2018Bonjean, 2022Hincks, 2024Isopi}.

The standard approach to measuring the tSZ effect involves combining CMB temperature maps at multiple frequencies into a single Compton-$y$ map, which provides a dimensionless measure of the integrated pressure along each line-of-sight. This is typically done with an internal linear combination (ILC) of the individual-frequency maps that isolates each of the signals with a unique spectrum into its own map: the CMB plus the kSZ signal, the tSZ signal, and the cosmic infrared background (CIB; see, e.g., \citealt{2024Coulton, 2020Madhavacheril}).  Compton-$y$ maps have been used to study the physical properties of galaxy clusters and filaments (e.g., \citealt{2013PlanckFil, 2018Bonjean, 2024Coulton, 2024Isopi}). Since the kSZ signal and the primary CMB have the same spectral dependence, as discussed above, the kSZ effect is not present in Compton-$y$ maps. Furthermore, accounting for relativistic corrections in a Compton-$y$ map requires making \textit{a priori} model assumptions about the gas temperature (see, e.g., \citealt{2024bCoulton}).

In this paper, we present a framework to model the SZ signal in galaxy clusters and filaments using a multifrequency and multi-instrument approach on a per-pixel basis. The advantage of this technique is that it can effectively separate the tSZ and the kSZ effects, while also accounting for relativistic corrections. In addition, this methodology allows for flexible per-pixel spatial and spectral modelling of astrophysical foregrounds, such as point sources, dust emission, and radio sources. Our approach is modular and enables the integration of new data from upcoming instruments across a wide range of frequencies. 

Our pipeline was first applied in \citet{2023Marrewijk}, who performed joint model fitting of ACT and ALMA data of the protocluster XLSSC~122. However, the application of our technique in that paper was limited to exploring only the tSZ effect, using just two ACT maps (98 and 150\,GHz), and the methodology was only briefly described. In this paper, we provide a full description of the method and demonstrate its capability to jointly model multiple astrophysical signals across a wide range of frequencies. We demonstrate this technique by applying it to the Abell 399--401 galaxy cluster pair and filament system (A399--401) using data from the Atacama Cosmology Telescope (ACT) and the \textit{Planck} satellite. Table \ref{tab:info_a399_401} presents some basic information about the system.

\begin{table}
\centering
\begin{minipage}{\columnwidth} 
\centering
\begin{tabular}{|c|c|c|c|}
\hline
\textbf{Parameter}          & {\textbf{A401}}    & \textbf{A399}    & \textbf{Filament}         \\ \hline
RA                & {44.741$^{\circ}$}   & 44.464$^{\circ}$   & 44.661$^{\circ}$            \\ \hline
DEC               & {13.580$^{\circ}$}   & 13.040$^{\circ}$  & 13.322$^{\circ}$           \\ \hline
$z$                 &  0.07366  & 0.07181    & 0.07274             \\ \hline
Separation & \multicolumn{2}{c|}{36.21$^{\prime}$ (3.11 Mpc)} & -- \\ \hline
Length            & --    & --      & 15.6$^\prime$   \\ \hline
Width             &   --   & --       & 12.3$^\prime$   \\ \hline
\end{tabular}
\caption{Basic information about the Abell 399--401 system. The redshifts are taken from \citet{2001OegerleHill}. We take the filament redshift to be the average redshift between the two clusters. All other parameters are the best-fit values calculated in this work, as shown in Section~\ref{sec:results}.}
\label{tab:info_a399_401}
\end{minipage}
\end{table}


\section{Instrument and data overview}
\subsection{Atacama Cosmology Telescope}\label{ssec:act}
ACT was a six-meter telescope located at an altitude of 5190 meters in the Atacama Desert in Chile that observed in bands roughly centered at 30, 40, 98, 150, and 220\,GHz; see Figure~\ref{fig:bandpasses_mfreq} for plots of the bandpasses. The telescope had a beam full-width-at-maximum (FWHM) of 2.1$^{\prime}$ at 98\,GHz and 1.4$^{\prime}$ at 150\,GHz. The Atacama Desert is a dry site with low precipitable water vapor, which makes it an excellent location for microwave astronomy \citep{2022Morris, 2025Morris}. The ACT project went through three versions of detectors: (i) the Millimeter Bolometric Array Camera (MBAC), which collected data from 2007 to 2010 \citep{fowler2007, 2011ApJSSwetz} (ii) ACTPol, which observed both the CMB temperature and polarization data from 2013 to 2015 \citep{Thornton2016}, and (iii) Advanced ACTPol, which collected temperature and polarization data from 2016 to 2022 \citep{2016Henderson}.

A399--401 is located within the regular ACT survey area, but to better probe this unique system, 31.5 hours of targeted observations were obtained with ACT in November 2020. This roughly doubled the ACT map depth of the system. We combined these dedicated observations with the regular DR6v2 survey data collected between 2017 and 2020. These data include observations from three ACT polarized arrays (PA) of detectors, PA4, PA5, and PA6, which provided coverage at 98, 150, and 220\,GHz. In this analysis, we consider a $2.1^\circ \times 2.1^\circ$ region centered on A399--401, corresponding to 252 pixels on each side with a $0.5'$ pixel scale. While more recent ACT DR6 data are available \citep{act_dr6_maps}, we did not incorporate them in this analysis. The targeted observations remain the dominant contributor to the high sensitivity in our maps, and the inclusion of newer data would offer only marginal improvements while requiring substantial additional effort in handling associated map-based systematics. Although the multifrequency methodology we developed in this work allows for fitting individual point sources, we used source-subtracted maps for this analysis to reduce the number of fit parameters. Future analyses may benefit from explicitly modelling the point sources in the fit. Finally, although ACT also observed the system at 30 and 40\,GHz, the data processing and calibration of these channels is still in process and not yet mature enough for inclusion in most scientific work. 

\begin{figure*}
    \centering
    \includegraphics[width=.8\textwidth]{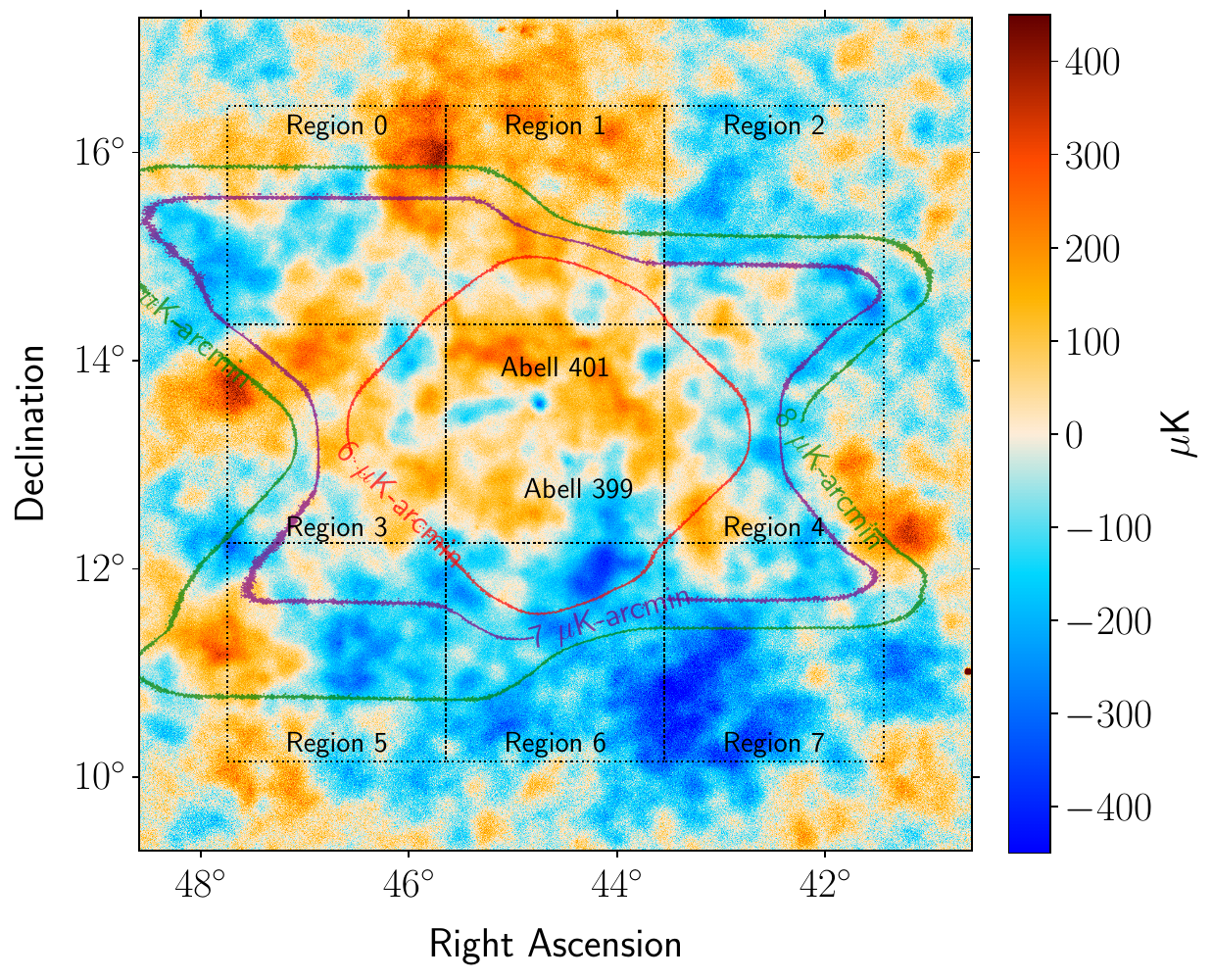}
    \caption{A coadded temperature map of ACT 98\,GHz PA5 showing the A399--A401 cluster pair and regions around the pair used for estimating the signal and noise covariance. The sensitivity contours are from the combined coadded variance maps from ACT and show that the data are deeper towards the center due to the 31.5 hours of dedicated observations of this system.}
    \label{fig:regions_map}
\end{figure*}

Four `splits' were made of each temperature map: the raw time-ordered data were sorted into single-day blocks, which were then optimally distributed across four independent sets to ensure even data distribution and uniform sky coverage \citep{2020Aiola, 2020JCAChoi, act_dr6_maps}, and made into separate maps, or `splits'. The splits should contain the same astrophysical signal but have independent noise realizations, which can be used to estimate their noise properties. We used temperature maps and did not require the use of polarization information. Figure~\ref{fig:regions_map} shows the central fit region of A399--A401 and the surrounding regions used for covariance estimation on the ACT PA5 98\,GHz coadd map. The sensitivity contours from the combined ACT maps are also shown, highlighting the higher sensitivity towards the center due to the 31.5 hours of targeted observations. The non-uniform depth of the maps is discussed further in Section~\ref{ssec:cov_variation}.

\subsection{Planck satellite}\label{ssec:planck}
The \textit{Planck} mission was a 1.5-meter telescope that observed the CMB from 30\,GHz to 857\,GHz, with resolutions ranging from $\sim 5^{\prime}$ to $32^{\prime}$. It measured the CMB anisotropies over the entire sky from 2009 to 2013, providing cosmological parameter constraints consistent with the $\Lambda$CDM model \citep{planck2018}. \textit{Planck} carried two instruments: the Low Frequency Instrument (LFI), with bands centered at approximately 30, 40, and 70\,GHz, and the High Frequency Instrument (HFI), with approximate band centers of 100, 143, 210, 353, 545, and 857\,GHz. In this work, we used point-source subtracted \textit{Planck} temperature maps from the \texttt{npipe} release of the LFI and HFI. We did not include the 857\,GHz channel due to its uncertain calibration \citep{planck_2016}. Two splits are available for each temperature map for the \textit{Planck} \texttt{npipe} data, corresponding to half-ring splits for LFI and half mission splits for HFI. Each map is reprojected from HEALPix \citep{gorski2005} to a Plate Carr\'ee (CAR) projection \citep{2002Calabretta} with a 0.5\,arcmin pixel size.

\subsection{X-ray imaging data}
We compared our best-fit results with the \textit{Chandra} broad band (0.5-7.0 keV) image of the A399--A401 system. We constructed the map from 24 Advanced CCD Imaging Spectrometer (ACIS) observations, which included both ACIS-I and ACIS-S observations that span over two decades of \textit{Chandra} operations, resulting in non-uniform coverage of the A399--401 field.\footnote{The following \textit{Chandra} observation identification numbers (ObsIDs) were used: 518, 3230, 2309, 10416, 10417, 10418, 10419, 14024, 22635, 22882, 26931, 27198, 27199, 27200, 27201, 27549, 27560, 27561, 27562, and 27586.} Data were reprocessed using the standard Chandra Interactive Analysis of Observations (CIAO; \citealt{2006Fruscione}) package 4.17.0. The observations were then light-curve filtered for flare events using the  {\tt deflare} task with {\tt method=clean}, and the resulting clean intervals were merged into an exposure-corrected image using the {\tt merge\_obs} task. The final image after processing is binned $4\times4$ pixels for a resolution of $1.968''$ and comprises $\sim$452\,ks of exposure time. 

\subsection{LOw-Frequency ARray data}

We include the LOw-Frequency ARray (LOFAR) radio data from \cite{deJong2022} for comparison of the diffuse radio emission with the best-fit Compton-$y$ and \textit{Chandra} X-ray images.  These continuum radio interferometric observations span 120-168\.MHz and are binned to a central frequency of 144\,MHz.
Due to the wide field nature of these observations, \cite{deJong2022} used \texttt{WSClean} \citep{Offringa2014}, a modern, multi-faceted, multi-scale deconvolution algorithm related to the original \texttt{CLEAN} algorithm \citep{Hogbom1974}.  For more details on the reduction, we refer the reader to \cite{deJong2022}.

\section{Methodology}\label{sec:methods}
This section presents the methodology for the multifrequency approach. We discuss the spectral and spatial models for SZ and dust signal in Section \ref{ssec:models}. The treatment of bandpasses and instrument beams is discussed in Sections \ref{ssec:bandpasses} and \ref{ssec:beams}, respectively. The procedure for estimating the covariance is described in Section \ref{ssec:cov}. The likelihood estimation and the model fitting procedure is discussed in Section \ref{ssec:like}. We also compared our results with a Compton-$y$ map that we made from the same multifrequency maps.

\subsection{Models} \label{ssec:models}
The CMB temperature decrement due to the thermal and kinematic SZ effect is
\begin{equation}
    \Delta T_{\rm SZ} (x) = T_{\rm CMB} \int dl~ \Bigg[\frac{k_{\rm B} T_{\rm e}}{m_{\rm e} c^{2}} f(\nu, T_{\rm e}) -\frac{v_{\rm r}}{c}\Bigg] n_{\rm e} \sigma_{\rm T},
    \label{eqn:sz_effect}
\end{equation}
where $l$ is the distance along the line of sight, $T_{\rm e}$ is the electron temperature, $\nu$ is the frequency of the light, $v_{\rm r}$ is the peculiar velocity projected along the line of sight, and $n_{\rm e}$ is the electron number density; as usual, $k_{\rm B}$, $m_{\rm e}$, $c$ and $\sigma_{\rm T}$ are Boltzmann's constant, the mass of the electron, the speed of light and the Thomson cross section, respectively. The function $f(\nu, \theta_{\rm e})$ encodes the frequency dependence of the tSZ effect,
\begin{equation}
    f(\nu, T_{\rm e}) = A_{0}(\nu) + \delta(\nu, T_{\rm e}).
    \label{eqn:rsz_effect}
\end{equation}
Here, $A_0(\nu) = x\coth({x / 2}) - 4$, where $x \equiv h\nu / k_{\rm B} T_{\rm CMB}$, $h$ is Planck's constant and $T_{\rm CMB} = 2.726$\,K \citep{2009Fixen}; it represents the temperature-independent contribution to the tSZ spectrum, while $\delta(\nu, T_{\rm e})$ accounts for the temperature-dependent relativistic corrections (see, e.g., \citealt{1998Itoh, 2004Itoh, CNSN, 2025Remazeilles}). In the non-relativistic Kompaneets approximation, $f_{\rm NR}(\nu) = A_{0}(\nu)$. Relativistic corrections become important at temperatures pertinent to the intercluster medium ($\gtrsim$~1\,keV), and therefore we include these corrections in our modelling. For the kSZ effect, a negative (positive) radial velocity means that the cluster is moving towards (away from) the observer, resulting in a temperature increase (decrease) in the CMB.

We model the tSZ and kSZ signal on a per-pixel basis using the \texttt{SZpack} numerical library, which allows for fast and precise computation of SZ signals up to high electron temperatures ($\sim$25\,keV; \citealt{CNSN, CSNN}). For the spatial profile of the SZ signals in the clusters, we used the classical $\beta$-profile to model the electron density within galaxy clusters \citep{1976Cavaliere},
\begin{equation}
    n_{\rm e}(r) = \frac{n_{\rm e0}}{[1 + (r/r_{\rm c})^{2}]^{3 \beta / 2}},
\end{equation}
where $r$ is the distance from the center of the cluster, $r_{\rm c}$ is the core radius of the cluster, $n_{\rm e0}$ is the density at the cluster center, and $\beta$  determines the slope of the profile. Although we used the $\beta$ profile for this work in order to compare our results with \citealt{2022Hincks} (hereafter H22), our methodology is flexible and allows for using any other profile, such as the generalized Navarro, Frenk, and White (gNFW) profile \citep{1996Zhao, 1997NFW} introduced for pressure profile modelling by \cite{2007Nagai}.
We assume an isothermal gas temperature and use the {elliptical} $\beta$-model for the spatial profile of the gas, projected to two dimensions on the sky (see \citealt{Hughes1998} for the derivation of these equations). The total observed two-dimensional model of the SZ signal is of the form
\begin{equation}
I_{\rm{SZ, model}} = I_{\rm{SZ}} \, \Bigg[1 + \frac{x^{{\prime} ^{2}} + (e y^{\prime})^{2}}{r_{\rm c}^{2}} \Bigg]^{0.5 - 1.5 \beta},
\end{equation}
where
\begin{equation}
\begin{split}
    x^{\prime} &= (x - x_{0}) \cos{\theta} - (y - y_{0}) \sin{\theta},\\
    y^{\prime} &= (x - x_{0}) \sin{\theta} + (y + y_{0}) \cos{\theta},
\end{split}
\end{equation}
$x_{0}$ and $y_{0}$ are cluster center positions in equatorial coordinates, $\theta$ is the position angle (defined such that North is zero degrees, and the angle increases in the counter-clockwise direction towards East), and $e$ is the major-to-minor axis ratio. $I_{\rm SZ}$ is computed using \texttt{SZpack} with the following input parameters:

\begin{itemize}
    \item $\tau_{\rm e}$, the optical depth;
    \item $T_{\rm e}$, electron temperature in keV;
    \item $\beta_{\rm o}$, the peculiar velocity of the observer with respect to the CMB rest frame ($\beta_{\rm o} = v_{\rm o} / c$);
    \item $\mu_{\rm o}$, the direction cosine of the line of sight with respect to the observer's velocity;
    \item $\beta_{\rm c}$, the peculiar velocity of the cluster or filament  ($\beta_{\rm c} = v_{\rm c} / c$);
    \item $\mu_{\rm c}$, the direction cosine from the cluster to the observer.  
\end{itemize}

We used the \texttt{COMBO} mode in \texttt{SZpack} for computing the SZ signal, which uses a combination of asymptotic expansions and basis functions, described further in \citet{CSNN, CNSN}. This mode provides fast and precise computation of the SZ signal for temperatures below 75\,keV.   The observer velocity with respect to the CMB rest frame is known from measurements of the CMB dipole, which indicate that the Sun is moving with $\beta_{\rm o} = 1.241 \times 10^{-3}$ towards the galactic coordinates $\ell = 264.14^{\circ}$ and $b = 48.26^{\circ}$ (\citealt{1996Fixsen, 2002Fixsen}). While the observer velocity is known and fixed, it must still be accounted for in the kSZ measurements, so that the kSZ signal can be transformed from the cluster rest frame to the CMB rest frame and then to the observer's rest frame; see \citet{2005Chluba, CNSN} for details. The direction cosine, $\mu_{\rm o}$, is computed from the cluster location on the sky relative to the direction of motion of the Sun with respect to the CMB. 

For the spatial profile of the filament, we adopt the ``mesa" model used in H22:
\begin{equation}
g(l, w) = \frac{1}{1 + (l/l_{0})^{8} + (w/w_{0})^{8} },
\end{equation}
where $l$ is the length of the filament parallel to the line joining the center of the clusters, and $w$ is the width of the filament in the direction perpendicular to $l$. 

\begin{figure}
    \centering
    \includegraphics[width=\columnwidth]{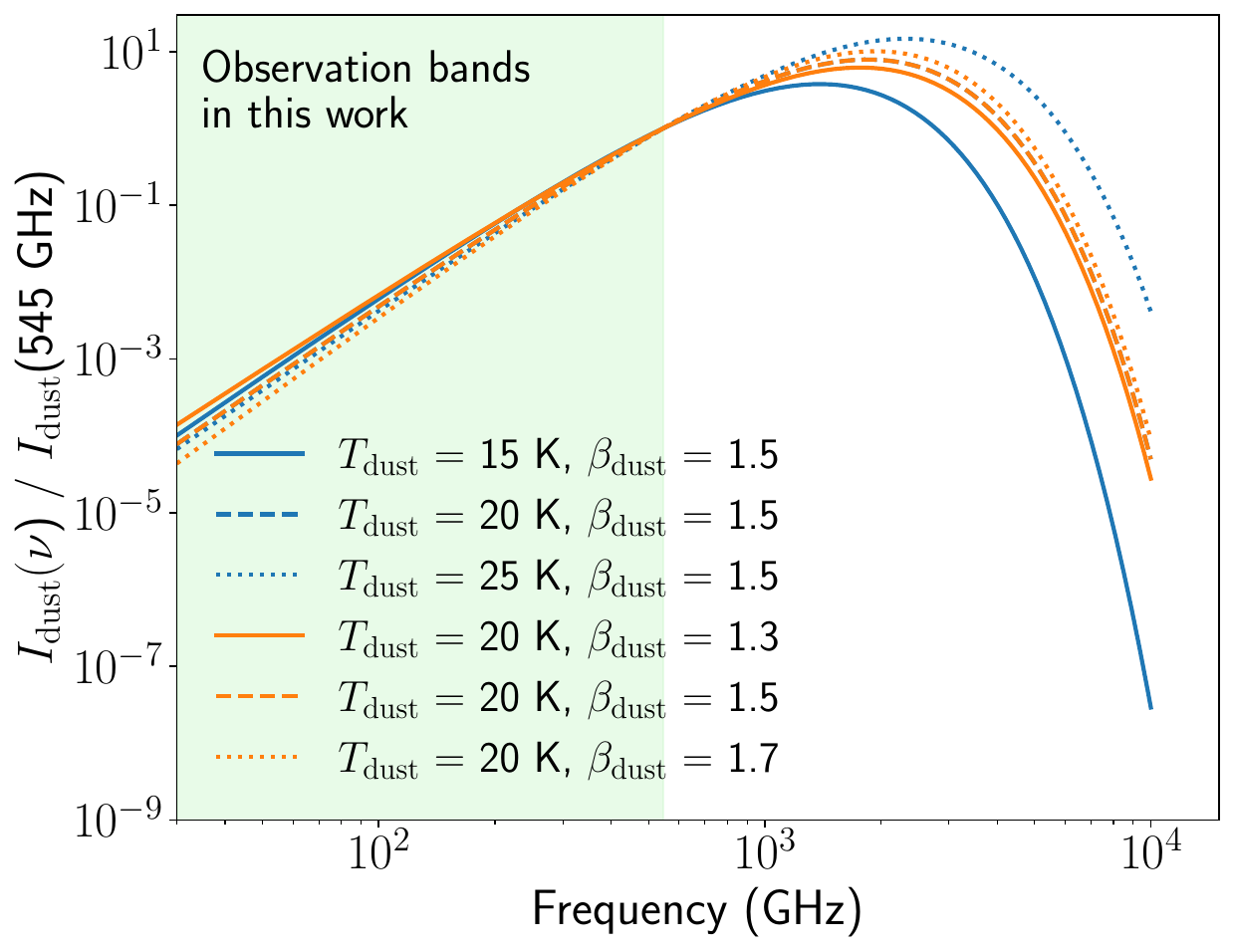}
    \caption{The modified blackbody spectrum from thermal dust emission in galaxy clusters (normalized to 545\,GHz) and the impact of variations in the dust temperature and emissivity index. The observation bands in this work lie in the Rayleigh-Jeans regime, so small changes in the dust temperature and emissivity index do not have a significant impact on the spectral shape.}
    \label{fig:dust_bands}
\end{figure}

To model the dust signal, we use a modified blackbody spectrum,
\begin{equation}
    I_{\rm{D}}  = A_{\rm{D}} \Bigg[\frac{\nu(1+z)}{\nu_{0}} \Bigg]^{\beta_{\rm{D}} + 3} \frac{\exp(h\nu_{0}/{k_{\rm B} T_{\rm D}})-1} {{\exp[h\nu(1+z)/{k_{\rm B} T_{\rm D}}]}-1},
\end{equation}
where $z$ is the redshift. We set the normalization frequency to $\nu_{0} = 545$\,GHz and fix the dust temperature, $T_{\rm D}$, to 20\,K and the dust emissivity index, $\beta_{\rm D}$, to 1.5. These values were selected based on the best-fit parameters from \citet{2016PlanckDust}, who fit the spectral energy distribution of dust using a stacked sample of more than 500 clusters observed with {\textit{Planck}} and the Infrared Astronomical Satellite. We allow the dust amplitude, $A_{\rm D}$, to be a free parameter. Our observation bands range from 30 to 545\,GHz, where the modified black body spectrum from thermal dust emission lies within the Rayleigh-Jeans (RJ) regime, where variations in the dust temperature and the dust emissivity index have a negligible impact on the dust spectral shape, as shown in Figure \ref{fig:dust_bands}.  We employ the same spatial profile for the dust signal as the SZ signal for the clusters and the filament. The total signal for the model is the sum of the SZ signal and the dust signal.

Note that we do not include the primary CMB as a separate model component. Although it shares its spectral shape with the kSZ effect, it acts as a large-scale background that does not follow the spatial profile of the clusters or the filament. This spatial distinction breaks the degeneracy, allowing us to account for it in our noise covariance treatment and measure the kSZ signal from the system.

The free parameters that we fit for each cluster are its position on the sky, the profile shape $\beta$, the core radius $r_{c}$, the ellipticity $e$, the position angle $\theta$, the optical depth $\tau_{\rm e}$, the electron temperature $T_{\rm e}$, and the dust amplitude $A_{\rm D}$. For the filament, we fit for the length $l_{0}$, the width $w_{0}$, as well as $T_{\rm e}$, $\tau_{\rm e}$, and $A_{\rm D}$. We fixed the position angle of the filament to align with the axis joining the two clusters. We adopt Gaussian priors on the gas temperatures. For A399 and A401, we used the results from XMM-Newton spectral fits \citep{2004Sakelliou},

\begin{equation}
    \begin{split}
        k_{\mathrm B}T_{\rm A401} &= 8.47 \pm 0.25\;\mathrm{keV}\\
        k_{\mathrm B}T_{\rm A399} &= 7.23 \pm 0.19\;\mathrm{keV}
    \end{split}
\end{equation}

For the inter-cluster filament, we used the results from \textit{Suzaku} observations \citep{2017Akamatsu}.

\begin{equation}
  k_{\mathrm B}T_{\rm fil} \;=\; 6.52 \pm 0.35\;\mathrm{keV}
\end{equation}

To model the kSZ effect and the peculiar velocities, we considered three cases:

\begin{enumerate}
    \item \textbf{Case~1: {Individual velocities}}. We fit the peculiar velocities of the two clusters and the filament individually. This leads to three free parameters: $v_{\rm A399}$, $v_{\rm A401}$, and $v_{\rm  fil}$. 
    
    \item \textbf{Case~2: {Bound system, fixed velocity difference}}. We treat the A399--A401 pair and the filament as a single, gravitationally bound system:
    
    \begin{equation}
      \begin{split}
        v_{\rm  A401} &= v_{\rm  system} + (v_{\rm  diff} / 2)\\
        v_{\rm  A399} &= v_{\rm  system} - (v_{\rm  diff} / 2)\\
        v_{\rm  fil}  &= v_{\rm  system}.
      \end{split}
    \label{eqn:bound_system}
    \end{equation}
    
    The only free parameter in the fit is the system's bulk velocity, $v_{\rm system}$. We fix the relative velocity between the two clusters to $v_{\rm diff} = 520\,\si{\kilo\metre\per\second}$, as derived from their spectroscopic redshift difference \citep{2001OegerleHill}. Equation~\ref{eqn:bound_system} assumes that the velocities of the system are non-relativistic, which is verified through self-consistency.

    \item\textbf{Case~3: {Bound system, free velocity difference}}. Following Equation~\ref{eqn:bound_system}, we fit for a single velocity for the whole system, but we additionally include $v_{\rm diff}$ as a free parameter rather than fixing it using the spectroscopic data as done in Case~2.
\end{enumerate}

\noindent The list of fit parameters and priors for the three cases is shown in Table~\ref{tab:priors}. We used Gaussian priors on the temperatures and uniform priors on all other parameters.

\begin{table*}[ht]
\centering
\begin{tabular}{l c c c}
\hline
\textbf{Parameter} & \textbf{Prior} & \textbf{Units} & \textbf{Description} \\
\hline
\multicolumn{4}{c}{\textbf{Abell 401}} \\ \hline
RA               & Uniform(44.651, 44.851)               & $^\circ$       & Right Ascension                     \\
Dec              & Uniform(13.472, 13.672)               & $^\circ$       & Declination                    \\
$\beta$             & Uniform(0.1, 2.0)                     & —              & $\beta$-model index                   \\
$r_c$            & Uniform(0.1, 6.0)                     & $^{\prime}$         & Core radius            \\
$e$              & Uniform(0.2, 2.0)                     & —              & Ellipticity ($b/a$)  \\
$\theta$            & Uniform(90, 140)                      & $^\circ$       & Position angle       \\
$T_{\rm e}$            & Gaussian(8.47, 0.25)  & keV            & Gas temperature                     \\
$\tau_{\rm e}$     & Uniform(0, 0.02)                      & —              & Optical depth          \\
$A_{\rm D}$            & Uniform(0, 2000000)               & Jy\,sr$^{-1}$              & Dust amplitude                        \\
\hline
\multicolumn{4}{c}{\textbf{Abell 399}} \\ \hline
RA               & Uniform(44.373, 44.573)               & $^\circ$       & Right Ascension                     \\
Dec              & Uniform(12.930, 13.130)               & $^\circ$       & Declination                    \\
$\beta$             & Uniform(0.1, 2.0)                     & —              & $\beta$-model index                 \\
$r_c$            & Uniform(0.1, 10.0)                    & $^{\prime}$         & Core radius            \\
$e$              & Uniform(0.2, 3.0)                     & —              & Ellipticity ($b/a$)  \\
$\theta$            & Uniform(75, 170)                      & $^\circ$       & Position angle       \\
$T_{\rm e}$            & Gaussian(7.23, 0.19)  & keV            & Gas temperature                     \\
$\tau_{\rm e}$     & Uniform(0, 0.02)                      & —              & Optical depth          \\
$A_{\rm D}$            & Uniform(0, 1000000)               & Jy\,sr$^{-1}$              & Dust amplitude                        \\
\hline
\multicolumn{4}{c}{\textbf{Filament}} \\ \hline
RA               & Uniform(44.60, 44.80)                   & $^\circ$       & Right Ascension                     \\
Dec              & Uniform(13.25, 14.00)                  & $^\circ$       & Declination                    \\
$l_0$            & Uniform(0, 30)                        & $^{\prime}$         &  Length                     \\
$w_0$            & Uniform(0, 30)                        & $^{\prime}$         &  Width                      \\
$T_{\rm e}$            & Gaussian(6.52, 0.35)   & keV            & Gas temperature                \\
$\tau_{\rm e}$     & Uniform(0, 0.005)                     & —              & Optical depth          \\
$A_{\rm D}$            & Uniform(0, 1000000)               & Jy\,sr$^{-1}$              & Dust amplitude                        \\
\hline
\multicolumn{4}{c}{\textbf{Velocity parameters: Case~1 (individual velocity fits)}} \\ \hline
$v_{\rm A401}$ & Uniform($-10000$, 10000)              & km\,s$^{-1}$   & Abell 401 peculiar velocity    \\
$v_{\rm A399}$ & Uniform($-10000$, 10000)              & km\,s$^{-1}$   & Abell 399 peculiar velocity    \\
$v_{\rm filament}$  & Uniform($-10000$, 10000)              & km\,s$^{-1}$   & Filament peculiar velocity \\
\hline

\multicolumn{4}{c}{\textbf{Velocity parameters: Case~2 (system velocity fit)}} \\ \hline
$v_{\rm system}$ & Uniform($-10000$, 10000)              & km\,s$^{-1}$   & A399-401 system velocity  \\
\hline

\multicolumn{4}{c}{\textbf{Velocity parameters: Case~3 (system velocity fit + pairwise velocity)}} \\ \hline
$v_{\rm system}$ & Uniform($-20000$, 20000)              & km\,s$^{-1}$   & A399-401 system velocity  \\
$v_{\rm diff}$ & Uniform($-20000$, 20000)              & km\,s$^{-1}$   & Velocity difference between A399 and A401  \\

\hline
\end{tabular}
\caption{Model parameters, priors, units, and descriptions.}
\label{tab:priors}
\end{table*}

\subsection{Bandpasses} \label{ssec:bandpasses}
\begin{figure}
    \centering
    \includegraphics[width=\columnwidth]{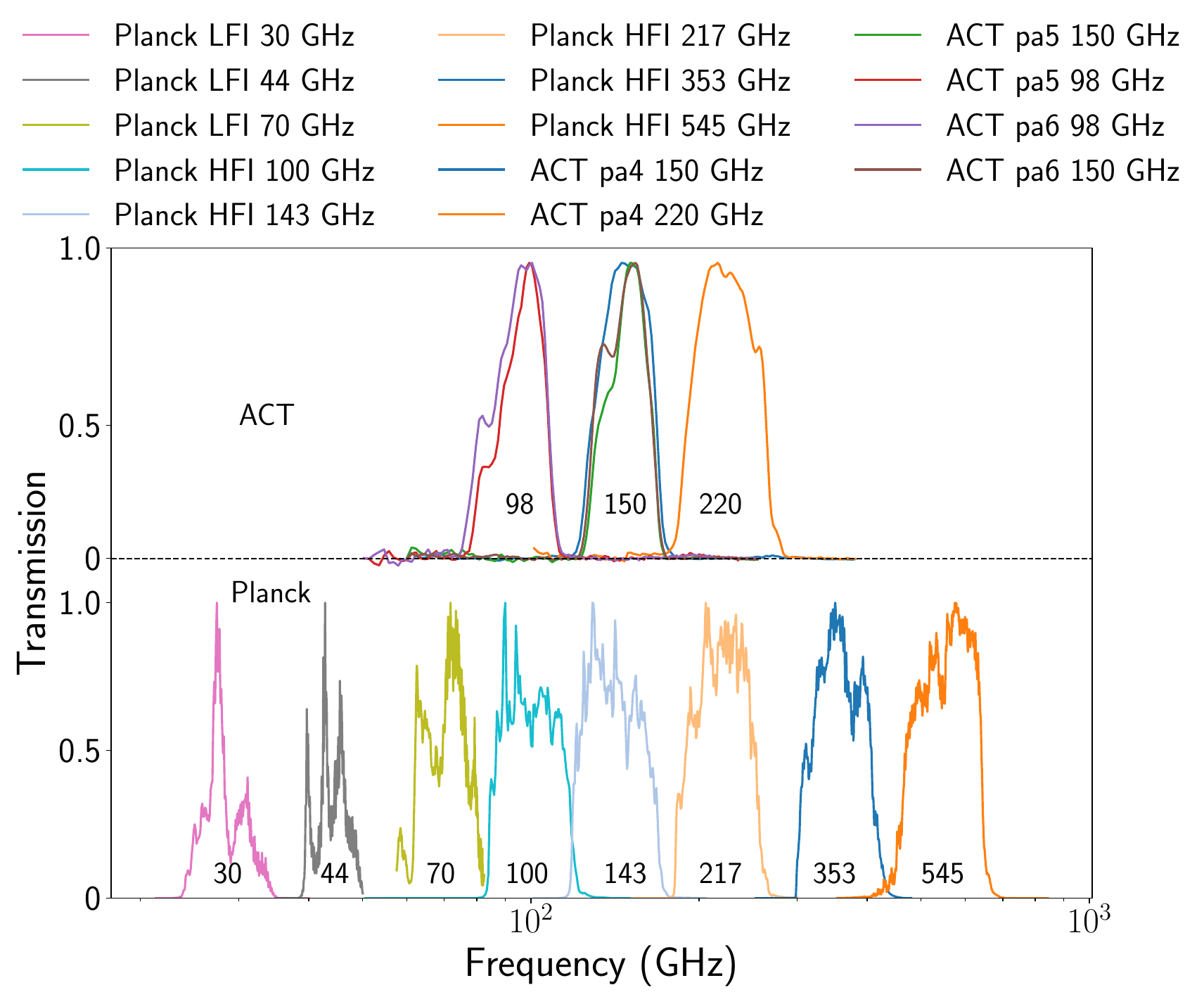}
    \caption{Bandpasses of the ACT and \textit{Planck} maps used in this work. These bandpasses have been normalized to have a peak of one.}
    \label{fig:bandpasses_mfreq}
\end{figure}

For a given array and frequency, we integrate the SZ intensity signal and the dust intensity signal over the normalized bandpass of the instrument. The bandpasses for ACT and \textit{Planck} are shown in Figure~\ref{fig:bandpasses_mfreq}. For \textit{Planck}, we used the LFI bandpasses described in \citet{2016PlanckBP} and HFI bandpasses described in \citet{2014PlanckBP}. For a given array and frequency, we integrate the signal as
\begin{equation}
    I_{\rm signal} =\frac{\int I_{\rm signal, \nu}\,R_{\nu}\,{\rm d}\nu}{\int R_{\nu}\,{\rm d}\nu},
\end{equation}
where $R_{\nu}$ is the bandpass and $I_{\rm signal, \nu}$ is the intensity and ``signal'' denotes either the SZ or the dust.

\subsection{Instrument beams} \label{ssec:beams}
We convolve the model with the instrument beams for ACT and \textit{Planck}. The ACT beams are measured from observations of Uranus and are approximated as being azimuthally symmetric  \citep{2022Lungu}. We model the \textit{Planck} beam as a two-dimensional Gaussian distribution with FWHM values taken from \citet{2016PlanckBeams}, which are listed in Table~\ref{tab:beams_act_planck}.

\begin{table}
\begin{tabular}{|c|c|c|}
\hline
\textbf{Instrument} & \textbf{Frequency} & \textbf{FWHM} \\ 
 & \textbf{(GHz)} & \textbf{(arcmin)} \\ 
\hline
\emph{Planck} &  30  & 32.4 \\ \hline
\emph{Planck} &  44  & 27.1 \\ \hline
\emph{Planck} &  70  & 13.3 \\ \hline
ACT           &  98  & 2.1  \\ \hline
\emph{Planck} & 100  &  9.7 \\ \hline
\emph{Planck} & 143  &  7.3 \\ \hline
ACT           & 150  & 1.4  \\ \hline
\emph{Planck} & 217  &  5.0 \\ \hline
ACT           & 220  & 1.0  \\ \hline
\emph{Planck} & 353  &  4.9 \\ \hline
\emph{Planck} & 545  &  4.6 \\ \hline
\end{tabular}
\caption{Summary of beam full-width at half-maximum (FWHM) for ACT and \emph{Planck}.}
\label{tab:beams_act_planck}
\end{table}


\subsection{Covariance} \label{sec:cov_est} \label{ssec:cov}
In our multifrequency framework, we simultaneously fit maps from multiple arrays, frequency bands, and instruments, in this case, ACT and \textit{Planck}. It is therefore necessary to account for the noise present in each individual map, but also the noise that is covariant between different maps. Here, by ``noise'' we mean any contribution to the map that is not captured by our model; hence, for instance, the primary CMB anisotropies are considered noise, as well as contributions more traditionally thought of in these terms, such as instrumental noise. Bearing this in mind, we separate the covariance into a linear combination of ``signal'', $S$, and ``noise'', $N$, components. The signal component includes the contribution from all the un-modelled components that are constant in time, including the CMB and astrophysical foregrounds. The noise component includes all time-varying signals, chiefly instrumental noise and atmospheric emission, and is calculated from the splits described in Sections~\ref{ssec:act} and \ref{ssec:planck}. The procedure for the covariance estimation broadly follows the procedure discussed in \cite{2020Madhavacheril} and is described further in this section.

\begin{figure*}
    \centering
    \includegraphics[width=0.7\linewidth]{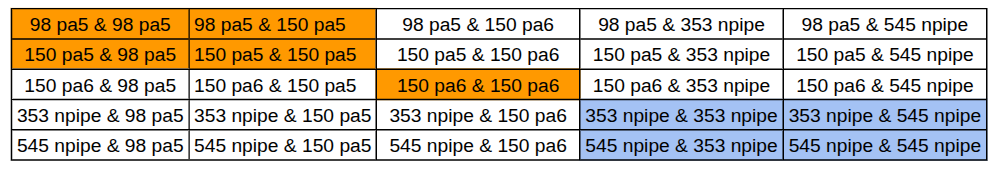}
    \caption{An illustrative schematic of part of the covariance matrix required for the multifrequency approach in this work. We show the portion of the covariance matrix that includes maps from ACT PA5 (98, 150\,GHz), ACT PA6 (150\,GHz), and \textit{Planck} (353 and 545\,GHz). The orange and blue elements are cases for which covariant noise must be included, whereas the signal covariance is computed for all matrix elements. The noise covariance is assumed to be zero for the matrix elements in white.}
    \label{fig:cov_matrix}
\end{figure*}

We expect the signal covariance to be present between all pairs of maps due to the contribution of the CMB signal and astrophysical foregrounds at all frequencies from 30 to 545\,GHz. For the {ACT} maps, we expect non-negligible noise covariance between frequency maps from the same array due to the detectors observing the same atmosphere and the presence of correlated instrumental noise on each array; we assume the noise covariance to be zero for all other cases. We also consider noise covariance between the \textit{Planck} channels to allow for potentially correlated noise between the channels. Figure~\ref{fig:cov_matrix} is a schematic representation of a small portion of the full covariance matrix, showing, by way of example, which combinations of ACT PA5 (98, 150\,GHz), ACT PA6 (150\,GHz), and \textit{Planck} (353 and 545\,GHz) maps require noise covariance estimates in addition to the signal covariance estimates. The total covariance for a given map pair is two-dimensional with pixel dimensions that match the fit region, which is $2.1^{\circ} \times 2.1^{\circ}$ or $252 \times 252$ pixels. 

\subsubsection{Instrument and atmospheric noise power spectrum}\label{sssec:noise_power}
This section describes the procedure for calculating the noise covariance for a given array, frequency, and instrument pair. 

\begin{enumerate}
    \item For a pair of array and frequency maps, we first define the notation array 1 and frequency 1 ($af$) and array 2 and frequency 2 ($a^{\prime}f^{\prime}$). For example, for the combination of ACT PA5 at 150\,GHz and ACT PA6 at 150\,GHz, we have $af$ = PA5--150 and $a^{\prime}f^{\prime}$ = PA6--150, respectively. For each array-frequency combination, we first compute the difference map for each split,
    
    \begin{equation}
        \mathbf{d}_{i} = \mathbf{s}_{i} - \mathbf{c},
    \end{equation}
    
where $\mathbf{c}$ is the coadd map ($\mathbf{c} = \sum \mathbf{s}_{i} \mathbf{h}_{i} / \sum \mathbf{h}_{i}$), $i$ is the index for different split maps ($1 \le i \le k$, where $k$ is the number of splits), and $\mathbf{h}_{i}$ is the per-pixel inverse variance map for each split. Subtracting a coadd map from the split map removes all the signal components, including the CMB, astrophysical foregrounds, as well as common systematics that persist across splits, such as ground pickup.

\item We then create a normalized inverse variance map, $\mathbf{h}_{i,n} = \mathbf{h}_{i} / \sum_{i} \mathbf{h}_{i}$.

\item We weight each difference map by $\mathbf{h}_{i,n}$,

\begin{equation}
\mathbf{d}_{i, w} = \mathbf{d}_{i} \times \mathbf{h}_{i,n}
\end{equation}

\item We then apply an apodized mask, $\mathbf{m}_{\rm{apod}}$, to eliminate edge effects when performing the Fast Fourier Transform (FFT), using a cosine profile that smoothly transitions from 1 to 0 at the map edges. We used a window width of $10'$ (20 pixels). 
 
\begin{equation}
    \mathbf{d}_{i, \text{apod}} = \mathbf{d}_{i, w} \times \mathbf{m}_{\rm{apod}}
\end{equation}

\item We then take the two-dimensional FFT of the normalized and apodized difference map to get the {noise (instrument and atmospheric) power spectrum},

\begin{align*}
    \tilde{\mathbf{d}}_{i, af} = \mathcal{F}(\mathbf{d}_{i, \text{apod}})
\end{align*}

\noindent where $\mathcal{F}$ is the FFT operator, and a tilde above a variable indicates that it is in Fourier space.

\item We estimate the noise power spectrum from eight background regions around the region containing the system being modelled. The size of each background region matches the size of the fit region. For A399--A401, we used a square box with a side length of 2.1$^{\circ}$. This box size is chosen such that the cluster fit region lies within an area with near-uniform map depth (see Figure \ref{fig:regions_map}). We scale the white noise level estimated from the outer regions by the level calculated from the central region, which is deeper. This procedure of noise scaling is further discussed in Section \ref{ssec:cov_variation}.
    
\item For a given map region, the noise power spectrum is

\begin{equation}
    (\mathbf{N}_{af, a^{\prime}f^{\prime}})_{\text{reg}} = \frac{1}{k(k-1)} \sum_{i} \frac{1}{w_{i}} (\tilde{\mathbf{d}}_{i, af} \times \tilde{\mathbf{d}}_{i, a^{\prime}f^{\prime}}^{*}),
\end{equation}

where the weight $w_{i} = \langle[(\tilde{\mathbf{h}}_{i, af} \times \mathbf{m}_{\rm{apod}}) \times (\tilde{\mathbf{h}}_{i, a^{\prime}f^{\prime}} \times \mathbf{m}_{\rm{apod}})]\rangle$ accounts for the loss in power due to the apodized mask. The $k-1$ factor is used to convert the noise power spectrum from the difference maps into a noise power estimate of the coadd map under the assumption of uniform noise weight per split.
\end{enumerate}

   
\subsubsection{Signal power spectrum}
The signal covariance is estimated from the eight outer regions around the fit region as follows:
\begin{enumerate}
\item The coadded maps are apodized:
\begin{equation}
  \begin{split}
    \mathbf{m}_{af, \text{apod}} &= \mathbf{m}_{af} \times \mathbf{m}_{\rm apod}, \\ \mathbf{m}_{a^{\prime}f^{\prime}, \text{apod}} &= \mathbf{m}_{a^{\prime}f^{\prime}} \times \mathbf{m}_{\rm apod},
  \end{split}
\end{equation}

and the covariance is calculated, accounting for the power loss from the apodization:
\begin{equation} \label{eq:sps}
\begin{split}
  (\mathbf{S}_{af, a^{\prime}f^{\prime}})_{\text{reg}} &= \frac{\mathcal{F}(\mathbf{m}_{af, \text{apod}}) \times \mathcal{F}(\mathbf{m}_{a^{\prime}f^{\prime}, \text{apod}})^{*}} 
  {\langle \mathbf{m}_{\rm apod} \times \mathbf{m}_{\rm apod} \rangle} \\
  &- (\mathbf{N}_{af, a^{\prime}f^{\prime}})_{\text{reg}}.
\end{split}
\end{equation}
\item The noise power spectrum estimate, as determined in Section~\ref{sssec:noise_power}, is then subtracted from the result of Equation~\ref{eq:sps}.


\end{enumerate}

\subsubsection{Total covariance}\label{sssec:total_covariance}

Finally, we combine the signal and noise covariances:

\begin{enumerate}
    \item For each element of the covariance matrix (each array and frequency combination), the total covariance, \textbf{P}, is the sum of the total signal power spectrum and the total noise power spectrum:
    \begin{align*}
        \textbf{P} &\equiv \mathbf{\bar{P}}_{af, a^{\prime}f^{\prime}} = \mathbf{S}_{af, a^{\prime}f^{\prime}} + \mathbf{N}_{af, a^{\prime}f^{\prime}} \\ &= \mathbf{\bar{S}}_{af, a^{\prime}f^{\prime}, \text{regs}} + \mathbf{\bar{N}}_{af, a^{\prime}f^{\prime}, \text{regs}},
    \end{align*} 
    where $\mathbf{\bar{S}}_{af, a^{\prime}f^{\prime}, \text{regs}}$ and $\mathbf{\bar{N}}_{af, a^{\prime}f^{\prime}, \text{regs}}$ are the average over the signal and noise power spectra from the eight background regions around the cluster fit region. 
    
    \item We smooth the total noise covariance with a Gaussian kernel with a standard deviation of 1.25 pixels ($0.75'$) to mitigate biases due to random fluctuations. We validated the choice of the smoothing kernel size by injecting synthetic clusters in the ACT and \textit{Planck} maps. When varying the smoothing kernel between one to two pixels ($0.5'$ to $1'$), we consistently recovered the input parameters.
\end{enumerate}

\begin{figure}[tb!]
    \centering
    \begin{minipage}{0.49\textwidth}
        \centering
        \includegraphics[width=0.9\linewidth]{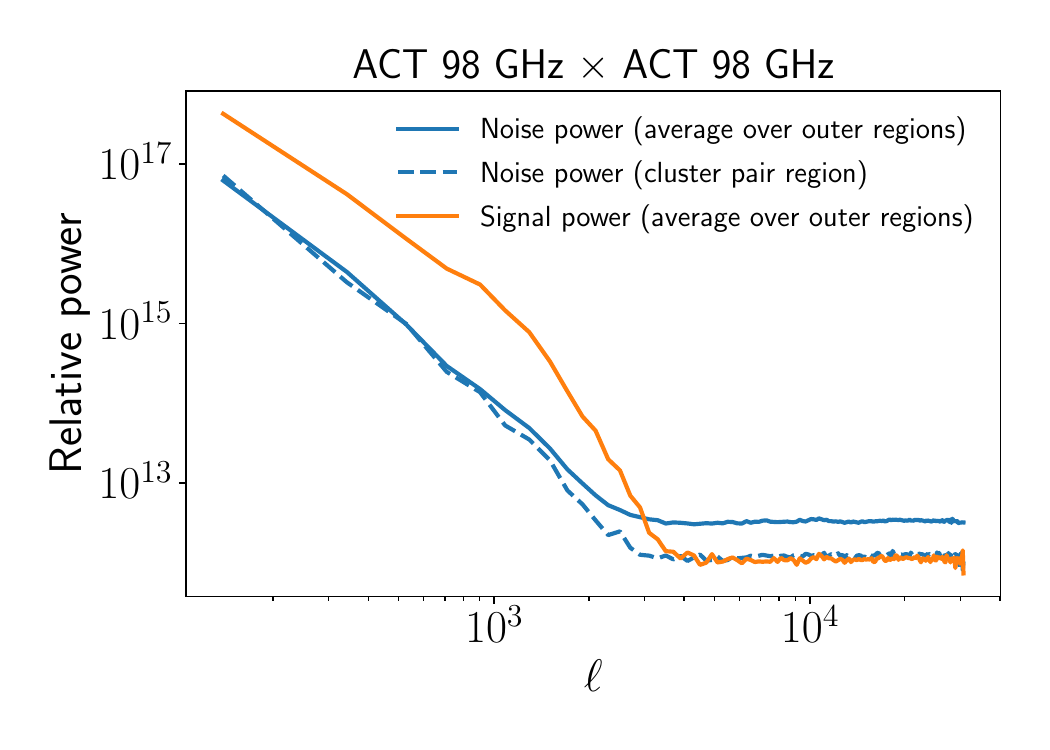}
    \end{minipage}
    \hfill
    \begin{minipage}{0.49\textwidth}
      \centering
      \includegraphics[width=.9\linewidth]{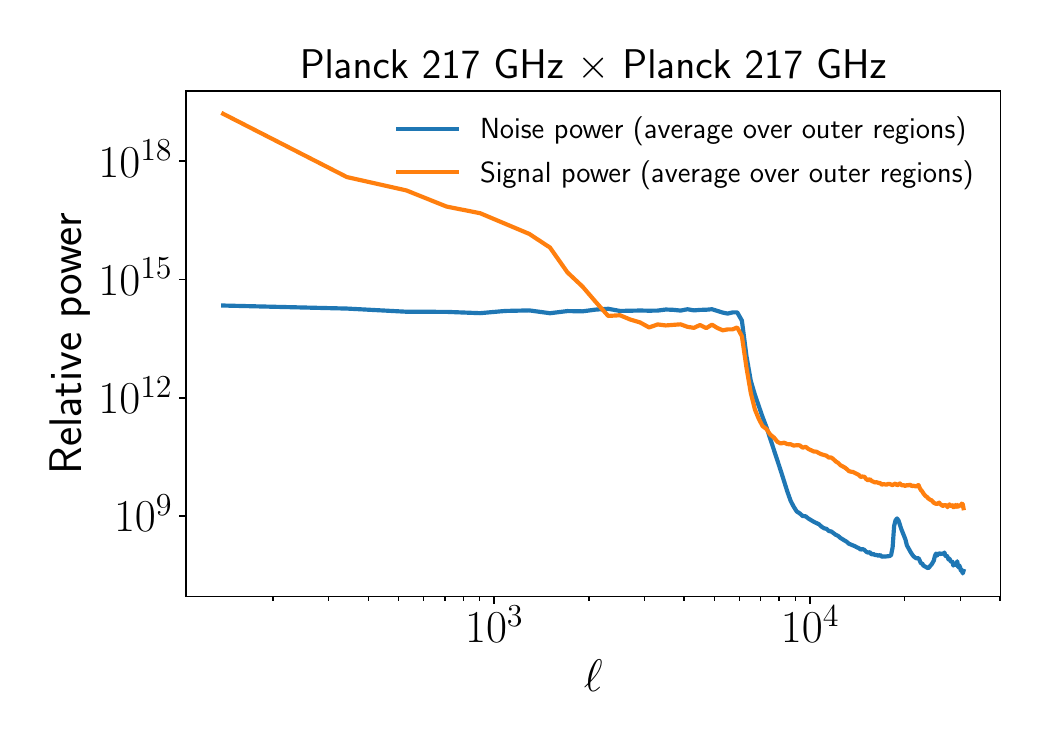}
    \end{minipage}
    \caption{The noise and signal power spectra as a function of multipole for ACT PA5 98\,GHz (top) and 217\,GHz \textit{Planck} (bottom). The noise levels in the cluster region for ACT are lower due to the deeper scan in that region as part of dedicated ACT observations of A399--401. The \textit{Planck} noise spectrum contains a knee at $\ell \sim 6100$ due to the finite band limit of the maps.}
    \label{fig:psd_combined}
\end{figure}

\begin{figure*}[htb!]
    \centering
    \includegraphics[width=0.75\textwidth]{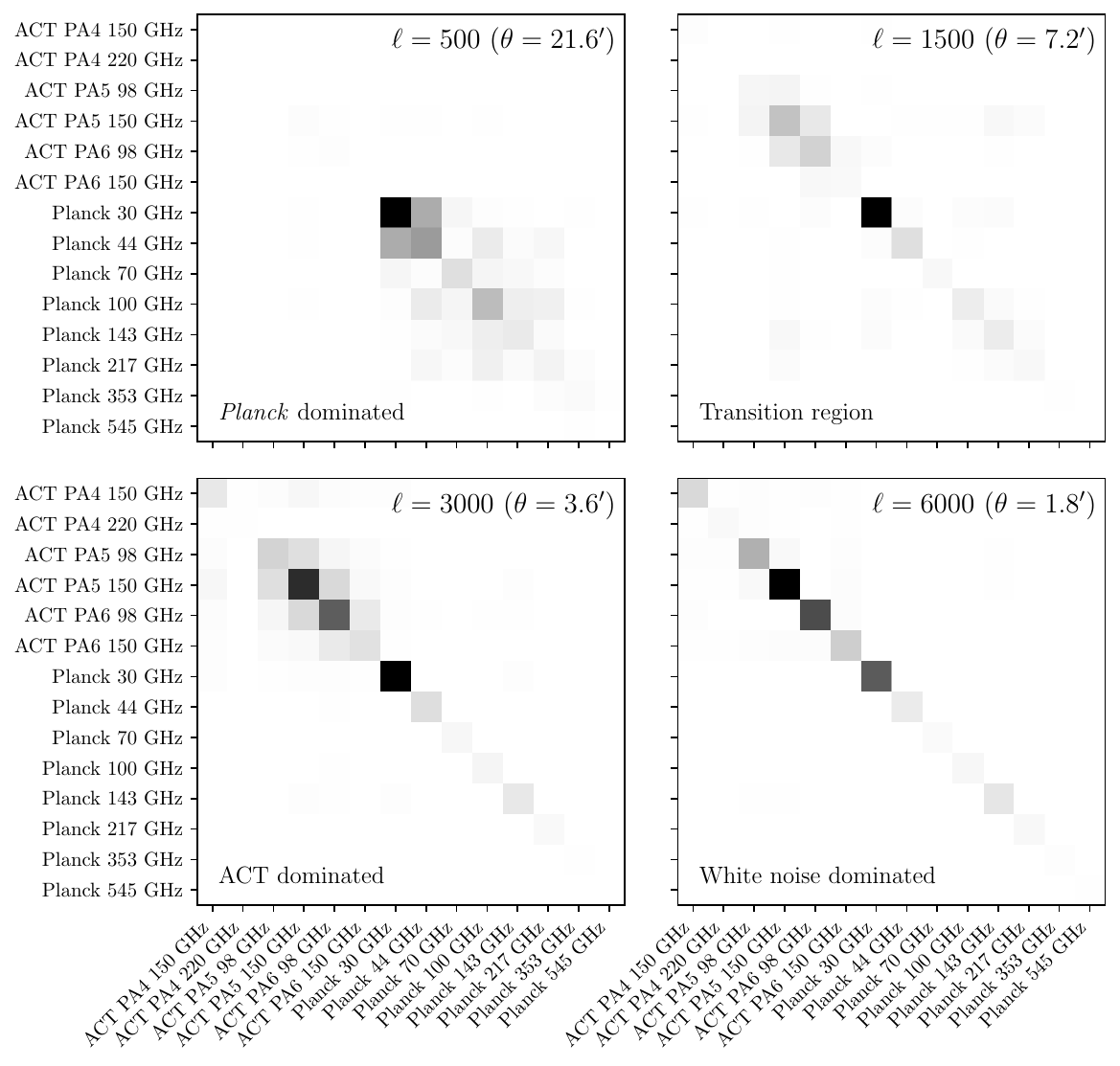}
    \caption{The relative statistical weights (calculated as inverse covariance) as a function of instrument, array, frequency, and angular scale. The maximum weight in each panel is normalized to 1 (shown as black) with linear scaling. This shows that the statistical power is dominated by \textit{Planck} on large scales ($\ell < 500$) and ACT on intermediate scales ($\ell > 1500$). On small scales ($\ell > 3000$), it approaches the white-noise dominated limit.}
    \label{fig:cov_vs_scale}
\end{figure*}

Figure~\ref{fig:psd_combined} shows the one-dimensional signal and noise power spectra of the ACT 98\,GHz PA5 map and the \textit{Planck} 217\,GHz map, by way of example. The noise from the CMB dominates at low $\ell$ (large angular scales) for the signal power spectrum. In the ACT noise power spectrum, the atmosphere dominates the noise at low $\ell$, and the detectors dominate the noise at high $\ell$ (small angular scales). The \textit{Planck} noise power spectrum is roughly constant and does not increase at low $\ell$ due to the lack of atmospheric noise. The \textit{Planck} noise spectrum contains a knee at $\ell \sim 6100$, which is due to finite band limit of the maps, as the maps were transformed to the CAR projection using a spherical harmonic transform $\ell_{\rm max} = 3 \times n_{\rm side}$, where $n_{\rm side} = 2048$. Figure~\ref{fig:cov_vs_scale} shows the total covariance as a function of instrument, array, frequency, and angular scale. This shows that the covariance is dominated by \textit{Planck} on large scales ($\ell \lesssim 500$), ACT on intermediate scales ($\ell \gtrsim 500$), and white noise on small scales ($\ell \gtrsim 3000$).

\subsubsection{Noise covariance estimate for A399--A401 with spatially varying depth} \label{ssec:cov_variation}
Due to the deeper ACT scans of the A399--A401 pair from 31.5 hours of targeted observations, the data are less noisy towards the center of the map, as shown by the dashed line in Figure~\ref{fig:regions_map}. Whereas the signal covariance should be approximately constant from region to region, the noise covariance will vary due to the non-uniform map depth. To model this spatial variability, we took the average of the noise covariance from the outer regions and then rescaled the white noise level at $\ell \geq 3000$ to match the white noise level measured from the central region. This method takes the slope of the atmospheric noise power as a function of scale to be constant, which is empirically found to be a reasonable approximation \citep[e.g.,][]{act_dr6_maps}, so that the benefit of longer observations is lowering the white noise level. However, we also tried modelling an $\ell$-dependent function from the central region and found that our results were not significantly different from rescaling by the white noise level.


\subsection{Model fitting procedure} \label{ssec:like}

To fit the model to our data, we seek to maximize the Gaussian likelihood:
\begin{equation}
    \mathcal{L} = \frac{1}{\sqrt{2\pi} |\mathbf{C}|^{1/2}} \exp{\left[-\frac{1}{2}(\mathbf{d} - \mathbf{m})^{T} \, \mathbf{C}^{-1} (\mathbf{d} - \mathbf{m})\right]},
    \label{eq:likelihood_general}
\end{equation}
where $\mathbf{d}$ is the vector of data, $\mathbf{m}$ is the vector of the model, and $\mathbf{C}$ is the covariance matrix. In our case, $\mathbf{d}$ consists of the maps from all arrays and frequencies from ACT and \textit{Planck}, which can be concatenated into a single vector,
%
\begin{equation}
    \mathbf{d} =
    \begin{bmatrix}
        \mathbf{d}_1 &
        \mathbf{d}_2 &
        \dots &
        \mathbf{d}_N
    \end{bmatrix}^T
\end{equation}
where $\mathbf{d}_{af}$ are the individual maps at each array--frequency combination $af$. The model $\mathbf{m}$ and covariance $\mathbf{C}$ can be constructed analogously. We  work in Fourier space, in which the residual maps, i.e., data minus model, are:
\begin{equation}
    \tilde{\mathbf{r}}_{af} = \frac{\mathcal{F}(\mathbf{d}_{af} \times \mathbf{m}_{\rm apod}) - \mathbf{\tilde{b}}_{af}\times\mathcal{F}(\mathbf{m}_{af} \times \mathbf{m}_{\rm apod})}{\langle \mathbf{m}_{\rm apod} \rangle},
    \label{eq:residual}
\end{equation}
where we apodize both the map and the model with an apodization window (10$^\prime$), $\mathbf{m}_{\rm apod}$, to eliminate any edge effects, and also convolve the model with the telescope beam window function, $\mathbf{\tilde{b}}_{af}$. In the denominator of Equation~\ref{eq:residual}, $\langle \mathbf{m}_{\rm apod} \rangle$ is the mean value of the apodization map, which properly renormalizes the result. Taking the logarithm of Equation~\ref{eq:likelihood_general}, ignoring its normalization (which is constant and does not matter for the optimization problem) and switching to component notation, we have:
\begin{equation}
    \log\mathcal{L} \propto -\sum_{af}\sum_{a'f'}\sum_k\sum_{k'} \tilde{r}_{af,k} \, \tilde{C}_{af,a'f',k,k'}^{-1} \, \tilde{r}_{a'f',k'}.
    \label{eq:loglike_full}
\end{equation}
where $k$ represents a Fourier pixel in a residual map $\mathbf{r}_{af}$. Our estimation of the covariances (Sec.~\ref{sec:cov_est}) implicitly assumes that the noise---whether `signal' noise or instrumental/atmospheric noise---is stationary, such that each individual covariance map is diagonal in Fourier space. Thus, the overall covariance is block-diagonal: $\tilde{C}_{af,a'f',k,k'} = \tilde{C}_{af,a'f',k,k'} \delta_{k,k'}$, and each Fourier pixel can be treated independently when computing the log-likelihood:
\begin{equation}
    \log\mathcal{L} \propto -\sum_k \left(\sum_{af}\sum_{a'f'} \tilde{r}_{af,k} \, \tilde{C}_{af,a'f',k}^{-1} \, \tilde{r}_{a'f',k}\right),
    \label{eq:loglike}
\end{equation}
where we have dropped the superfluous index $k'$ from $\tilde{C}$. The inversion of the covariance matrix in Equation~\ref{eq:loglike} is completely tractable because for each $k$, $\tilde{C}_{af,a'f',k}$ is only an $N \times N$ matrix, where $N = 14$ is the number of ACT and \textit{Planck} maps that we fit in this work.



We minimized Equation~\ref{eq:loglike} using the \texttt{emcee} \citep{Foreman-Mackey2013} Python package with the MCMC algorithm. We ran the MCMC for $>$50$\tau$ iterations, where $\tau$ is the autocorrelation time, to ensure that the chains have converged. We estimated the autocorrelation time by computing the autocorrelation function for each walker individually and then averaging these estimates, which leads to a more stable and lower variance estimate of $\tau$ compared to averaging all walkers into a single chain before computing the autocorrelation (see more in the documentation of \texttt{emcee}\footnote{https://emcee.readthedocs.io/en/stable/}). 







\section{Results} \label{sec:results}
 
\subsection{Best-fit Models}
\label{ssec:posterior_distributions}

\begin{figure}[htb!]
    \centering
    \begin{minipage}{0.49\textwidth}
        \centering
        \includegraphics[width=\linewidth]{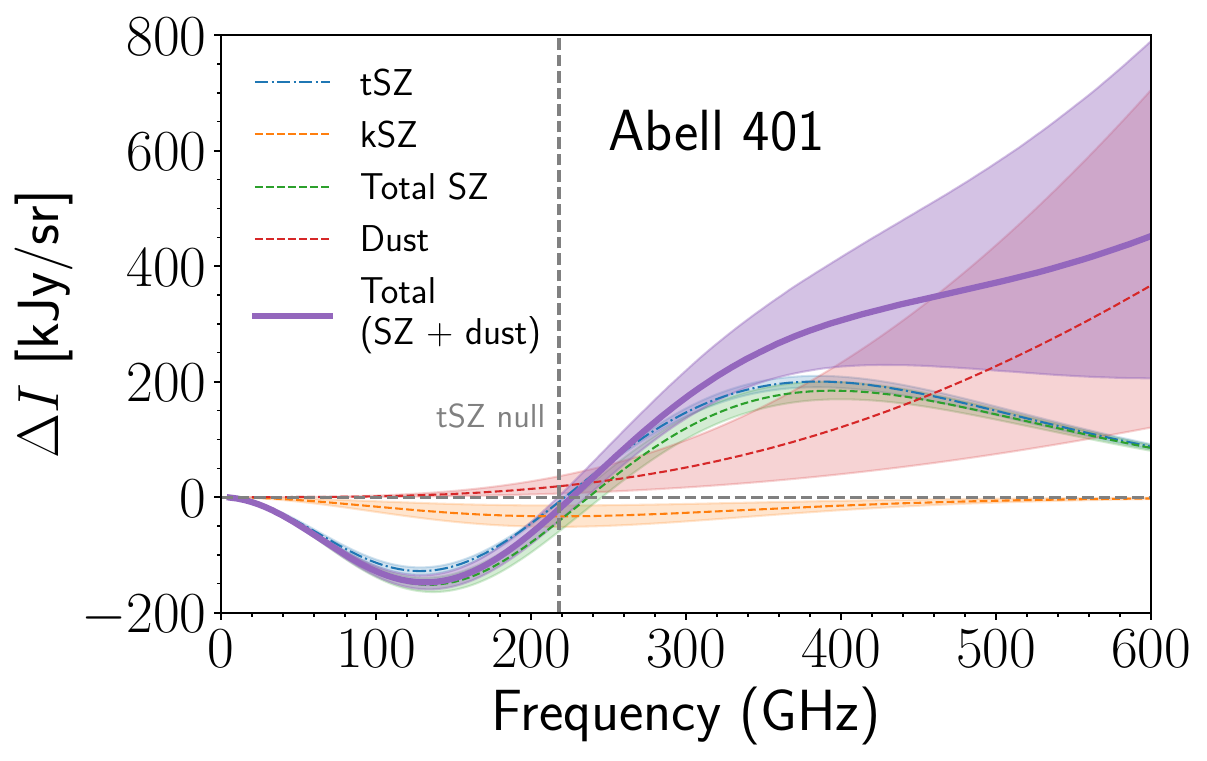}
    \end{minipage}
    \hfill
    \begin{minipage}{0.49\textwidth}
        \centering
        \includegraphics[width=\linewidth]{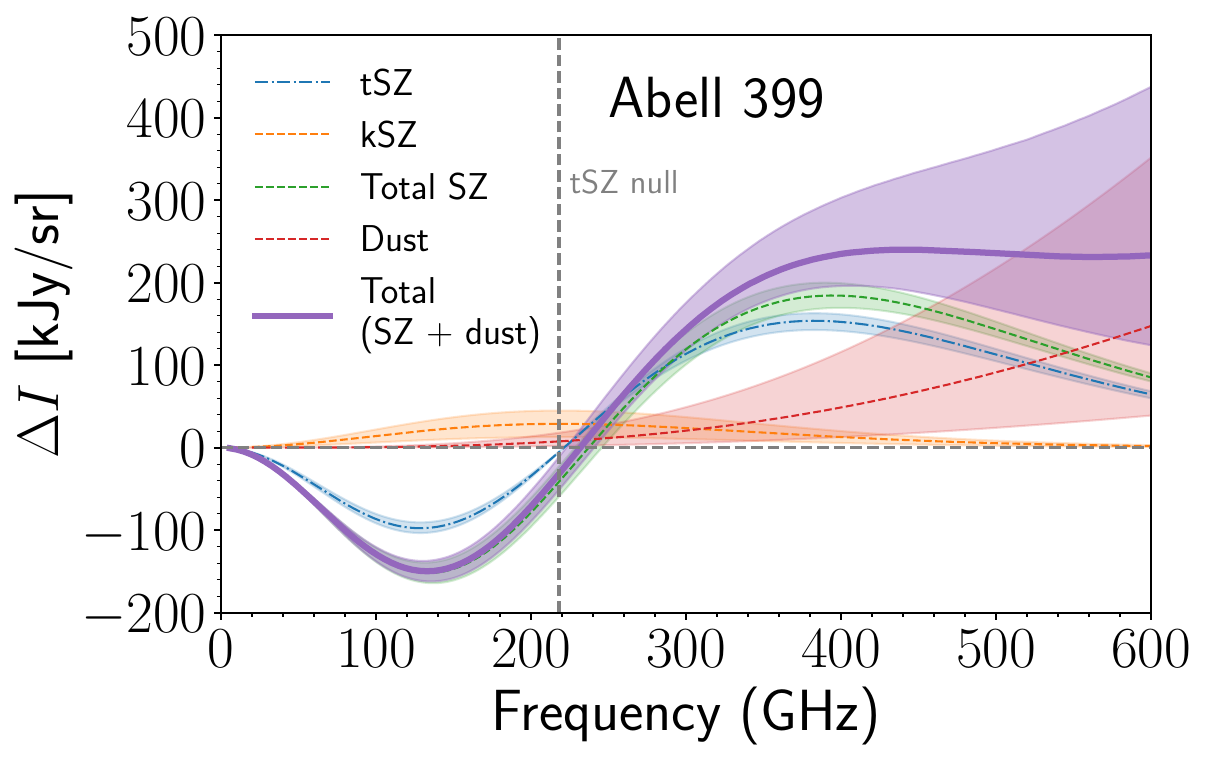}
    \end{minipage}
    \hfill
    \begin{minipage}{0.49\textwidth}
        \centering
        \includegraphics[width=\linewidth]{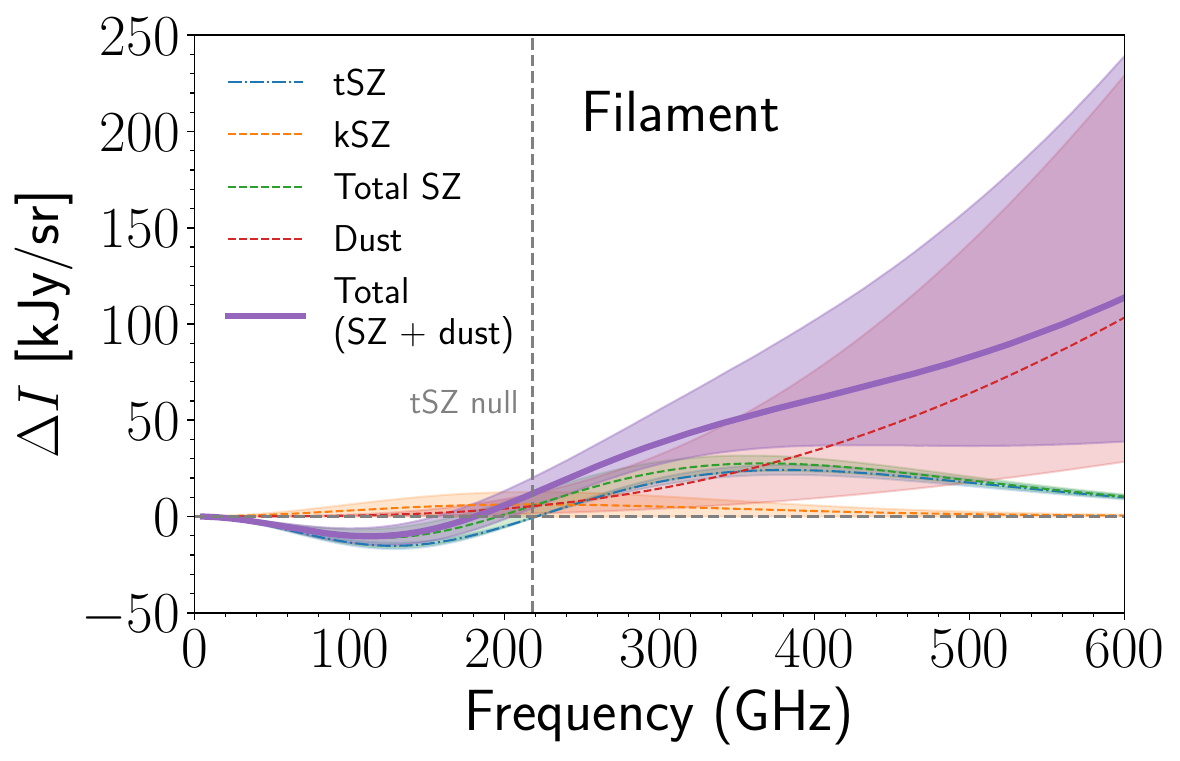}
    \end{minipage}
    \caption{The best-fit tSZ, kSZ, total SZ, dust, and total signal for the A399--A401 system for Case 1.}
    \label{fig:spectra}
\end{figure}

Table~\ref{tab:compare_table_y} shows the best-fit model parameters that we obtain for the A399--401 system, for each of the three kSZ cases described in Sec.~\ref{ssec:models}. We define the best-fit value as the median of the posterior distribution and report 1$\sigma$ error bars from the 16$^{\rm th}$ and 84$^{\rm th}$ percentiles. Fit results from H22 are also shown for comparison and are discussed further below. Corner plots of the posterior distributions are shown in the Appendix (Figures~\ref{fig:401_corner}--\ref{fig:v_cases2_and_3}). 

\begin{table*}[p] 
\centering
\footnotesize
\setlength{\tabcolsep}{3pt}
\begin{tabular}{|c|c|c|c|c|c|c|c|}
\hline
 \textbf{Parameter} & \textbf{Compton-$y$} & \textbf{Multifreq. (Case~1)} & \textbf{Multifreq. (Case~2)} & \textbf{Multifreq. (Case~3)} & \textbf{$|\Delta_{\rm ind}|$} & \textbf{$|\Delta_{\rm sys}|$} & \textbf{$|\Delta_{\rm diff}|$} \\
 & \citep{2022Hincks} & \textbf{(individual velocities)} & \textbf{(system velocity)} & \textbf{(system + diff)} & \textbf{($\sigma$)} & \textbf{($\sigma$)} & \textbf{($\sigma$)} \\ \hline

\multicolumn{7}{|c|}{\textbf{Abell 401}} \\ \hline
RA [$^{\circ}$]           & $44.751_{-0.002}^{+0.002}$        & $44.741_{-0.001}^{+0.001}$                  & $44.741_{-0.001}^{+0.001}$               & $44.741_{-0.001}^{+0.001}$                  & $4.47$  & $4.47$  & $4.47$  \\ \hline
DEC [$^{\circ}$]          & $13.572_{-0.002}^{+0.002}$        & $13.580_{-0.002}^{+0.002}$                  & $13.580_{-0.002}^{+0.002}$               & $13.580_{-0.002}^{+0.002}$                  & $2.83$  & $2.83$  & $2.83$  \\ \hline
$\beta$                   & $0.82_{-0.06}^{+0.07}$            & $1.112_{-0.056}^{+0.066}$                   & $1.094_{-0.054}^{+0.065}$                & $1.110_{-0.056}^{+0.066}$                   & $3.28$  & $3.11$  & $3.25$  \\ \hline
$r_{\rm c}$ [$^{\prime}$] & $2.6_{-0.3}^{+0.4}$               & $4.328_{-0.380}^{+0.410}$                   & $4.040_{-0.349}^{+0.379}$                & $4.272_{-0.370}^{+0.405}$                   & $3.27$  & $2.85$  & $3.20$  \\ \hline
$e$ ($b/a$)               & $0.81_{-0.05}^{+0.06}$            & $0.800_{-0.044}^{+0.046}$                   & $0.794_{-0.046}^{+0.050}$                & $0.800_{-0.046}^{+0.048}$                   & $0.14$  & $0.22$  & $0.14$  \\ \hline
$\theta$ [$^{\circ}$]     & $123_{-8}^{+9}$                   & $107.8_{-5.5}^{+5.7}$                       & $107.82_{-5.7}^{+5.7}$                    & $107.9_{-5.4}^{+5.6}$                       & $1.49$  & $1.48$  & $1.49$  \\ \hline
$y$                       & $(12.6_{-0.6}^{+0.6})\times10^{-5}$ & ($1.219 \pm 0.077)\times10^{-4}$ & ($1.258 \pm 0.079) \times10^{-4}$ & ($1.224 \pm 0.078)\times10^{-4}$ & $0.42$  & $0.02$  & $0.37$  \\ \hline
$T_{\rm e}$ [keV]         & --                                & $8.452_{-0.250}^{+0.249}$                   & $8.447_{-0.249}^{+0.254}$                & $8.455_{-0.249}^{+0.247}$                   & --       & --       & --       \\ \hline
$\tau_{\rm e}$            & --                                & $(7.368_{-0.411}^{+0.415})\times10^{-3}$ & $(7.612_{-0.409}^{+0.429})\times10^{-3}$ & $(7.398_{-0.415}^{+0.418})\times10^{-3}$ & --       & --       & --       \\ \hline
$A_{\rm D}$ [Jy/sr]       & --                                & $(2.37_{-1.59}^{+2.15})\times10^{5}$ & $(1.86_{-1.29}^{+1.95})\times10^{5}$ & $(2.28_{-1.53}^{+2.13})\times10^{5}$ & -- & -- & -- \\ \hline

\multicolumn{7}{|c|}{\textbf{Abell 399}} \\ \hline
RA [$^{\circ}$]           & $44.473_{-0.004}^{+0.004}$        & $44.464_{-0.003}^{+0.003}$                  & $44.464_{-0.003}^{+0.003}$               & $44.464_{-0.003}^{+0.003}$                  & $1.80$  & $1.80$  & $1.80$  \\ \hline
DEC [$^{\circ}$]          & $13.030_{-0.003}^{+0.003}$        & $13.040_{-0.003}^{+0.003}$                  & $13.040_{-0.003}^{+0.003}$               & $13.040_{-0.003}^{+0.003}$                  & $2.36$  & $2.36$  & $2.36$  \\ \hline
$\beta$                   & $0.81_{-0.08}^{+0.11}$            & $1.146_{-0.079}^{+0.107}$                   & $1.166_{-0.085}^{+0.135}$                & $1.147_{-0.080}^{+0.119}$                   & $2.53$  & $2.45$  & $2.45$  \\ \hline
$r_{\rm c}$ [$^{\prime}$] & $3.0_{-0.6}^{+0.7}$               & $4.343_{-0.516}^{+0.585}$                   & $4.533_{-0.524}^{+0.613}$                & $4.320_{-0.517}^{+0.601}$                   & $1.58$  & $1.78$  & $1.54$  \\ \hline
$e$ ($b/a$)               & $0.93_{-0.07}^{+0.05}$            & $0.934_{-0.075}^{+0.088}$                   & $0.921_{-0.077}^{+0.089}$                & $0.929_{-0.078}^{+0.091}$                   & $0.04$  & $0.09$  & $0.01$  \\ \hline
$\theta$ [$^{\circ}$]     & $133_{-26}^{+27}$                 & $107.2_{-25.3}^{+36.9}$                      & $109.6_{-19.8}^{+34.9}$                   & $111.1_{-21.7}^{+36.0}$                     & $0.63$  & $0.61$  & $0.56$  \\ \hline
$y$                       & $(8.1_{-0.6}^{+0.6})\times10^{-5}$ & $(9.215 \pm 0.699)\times10^{-5}$ & $(8.953 \pm 0.786) \times10^{-5}$ & $(9.211 \pm 0.738)\times10^{-5}$ & $1.21$  & $0.86$  & $1.17$  \\ \hline
$T_{\rm e}$ [keV]         & --                                & $7.228_{-0.191}^{+0.193}$                   & $7.214_{-0.193}^{+0.190}$                & $7.231_{-0.186}^{+0.186}$                   & --       & --       & --       \\ \hline
$\tau_{\rm e}$            & --                                & $(6.514_{-0.499}^{+0.427})\times10^{-3}$ & $(6.341_{-0.624}^{+0.437})\times10^{-3}$ & $(6.509_{-0.544}^{+0.443})\times10^{-3}$ & --       & --       & --       \\ \hline
$A_{\rm D}$ [Jy/sr]       & --                                & $(9.60_{-7.06}^{+13.36})\times10^{4}$        & $(1.08_{-0.78}^{+1.42})\times 10^{5}$     & $(9.42_{-6.90}^{+13.09})\times10^{4}$        & --       & --       & --       \\ \hline

\multicolumn{8}{|c|}{\textbf{Filament}} \\ \hline
RA [$^{\circ}$]           & $44.68_{-0.02}^{+0.02}$           & $44.661_{-0.024}^{+0.016}$                  & $44.658_{-0.036}^{+0.019}$               & $44.658_{-0.031}^{+0.017}$                  & $0.67$  & $0.65$  & $0.70$  \\ \hline
DEC [$^{\circ}$]          & $13.37_{-0.03}^{+0.02}$           & $13.322_{-0.028}^{+0.022}$                  & $13.322_{-0.048}^{+0.025}$               & $13.322_{-0.039}^{+0.023}$                  & $1.36$  & $1.08$  & $1.21$  \\ \hline
Length [$^{\prime}$]      & $12.3_{-1.4}^{+1.7}$              & $15.593_{-1.333}^{+2.542}$                  & $15.822_{-1.494}^{+6.341}$               & $15.674_{-1.395}^{+4.667}$                  & $1.33$  & $0.84$  & $0.99$  \\ \hline
Width [$^{\prime}$]       & $10.8_{-1.0}^{+1.1}$              & $12.270_{-0.907}^{+1.009}$                  & $12.413_{-1.056}^{+1.098}$               & $12.377_{-1.027}^{+1.116}$                  & $1.03$  & $1.07$  & $1.05$  \\ \hline
$y$                       & $(1.10_{-0.18}^{+0.17})\times10^{-5}$ & $(1.436 \pm 0.186)\times10^{-5}$ & ($1.471 \pm 0.187)\times10^{-5}$ & $(1.481 \pm 0.184)\times10^{-5}$ & $1.32$  & $1.45$  & $1.50$  \\ \hline
$T_{\rm e}$ [keV]         & --                                & $6.509_{-0.347}^{+0.347}$                   & $6.492_{-0.351}^{+0.358}$                & $6.483_{-0.345}^{+0.349}$                   & --       & --       & --       \\ \hline
$\tau_{\rm e}$            & --                                & $(1.127_{-0.130}^{+0.136})\times10^{-3}$ & $(1.116_{-0.130}^{+0.135})\times10^{-3}$ & $(1.167_{-0.127}^{+0.134})\times10^{-3}$ & --       & --       & --       \\ \hline
$A_{\rm D}$ [Jy/sr]       & --                                & $(6.61_{-4.76}^{+8.31})\times10^{4}$        & $(7.13_{-5.15}^{+9.08})\times10^{4}$     & $(7.50_{-5.39}^{+9.00})\times10^{4}$        & --       & --       & --       \\ \hline

\multicolumn{7}{|c|}{\textbf{Velocities: Case~1 (individual velocity fits)}} \\ \hline
$v_{\rm A401}$ [km/s]     & --                                & $+1026_{-580}^{+609}$                        & --                                & --                                    & --       & --       & --       \\ \hline
$v_{\rm A399}$ [km/s]     & --                                & $-1007_{-525}^{+553}$                         & --                                & --                                    & --       & --       & --       \\ \hline
$v_{\rm filament}$ [km/s] & --                                & $-1252_{-1361}^{+1290}$                        & --                                & --                                    & --       & --       & --       \\ \hline

\multicolumn{7}{|c|}{\textbf{Velocities: Case~2 (system velocity fits)}} \\ \hline
$v_{\rm system}$ [km/s]   & --                                & --                                & $-116_{-358}^{+376}$               & --                                    & --       & --       & --       \\ \hline

\multicolumn{7}{|c|}{\textbf{Velocities: Case~3 (system + velocity difference fits)}} \\ \hline
$v_{\rm system}$ [km/s]   & --                                & --                                & --                                & $-124_{-358}^{+366}$                  & --       & --       & --       \\ \hline
$v_{\rm diff}$ [km/s]     & --                                & --                                & --                                & $+1944_{-783}^{+786}$                & --       & --       & --       \\ \hline
\end{tabular}
\caption{Best-fit parameters for the Abell 401--Abell 399 galaxy cluster pair system from the application of the multifrequency methodology developed in this work. We compare our results to the Compton-$y$ results from \cite{2022Hincks}. The columns titled $\Delta_{\rm ind}$, $\Delta_{\rm sys}$, and $\Delta_{\rm diff}$ show the significance of the difference between the Compton-$y$ and the multifrequency results for the three kSZ cases where we considered: (i) Case~1: fit the individual velocities of Abell 399, Abell 401, and the filament, (ii) Case~2: fit one system velocity with a fixed velocity difference between the clusters, and (iii) Case~3: fit the system velocity and the velocity difference, respectively. We report the best-fit value as the median of the posterior distributions and uncertainties based on the 16$^{\rm th}$ and 84$^{\rm th}$ percentiles. The Compton-$y$ parameter in our multifrequency fitting results is derived from the optical depth and the temperature.
}
\label{tab:compare_table_y}
\end{table*}

\begin{figure*}[p]
    \centering
    \includegraphics[width=\textwidth]{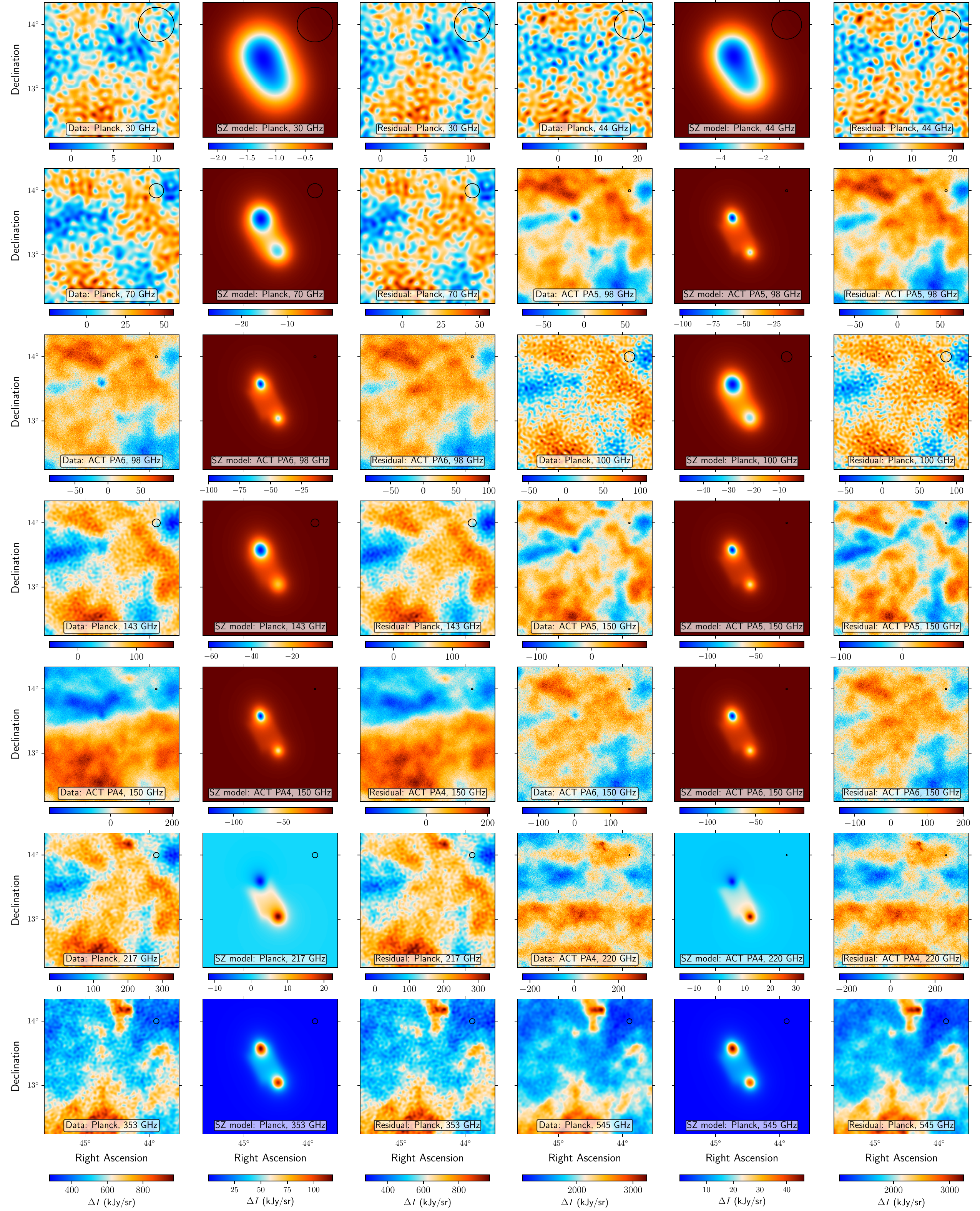}
    \caption{Data, best-fit SZ model, and residual plots from fitting A399--401 using our multifrequency approach. We used \textit{Planck} and ACT data from 30--545\,GHz. The beam size for each instrument and map is shown in the black circle.}
    \label{fig:bridge_30_545}
\end{figure*}

\begin{figure}
    \centering
    \includegraphics[width=\columnwidth]{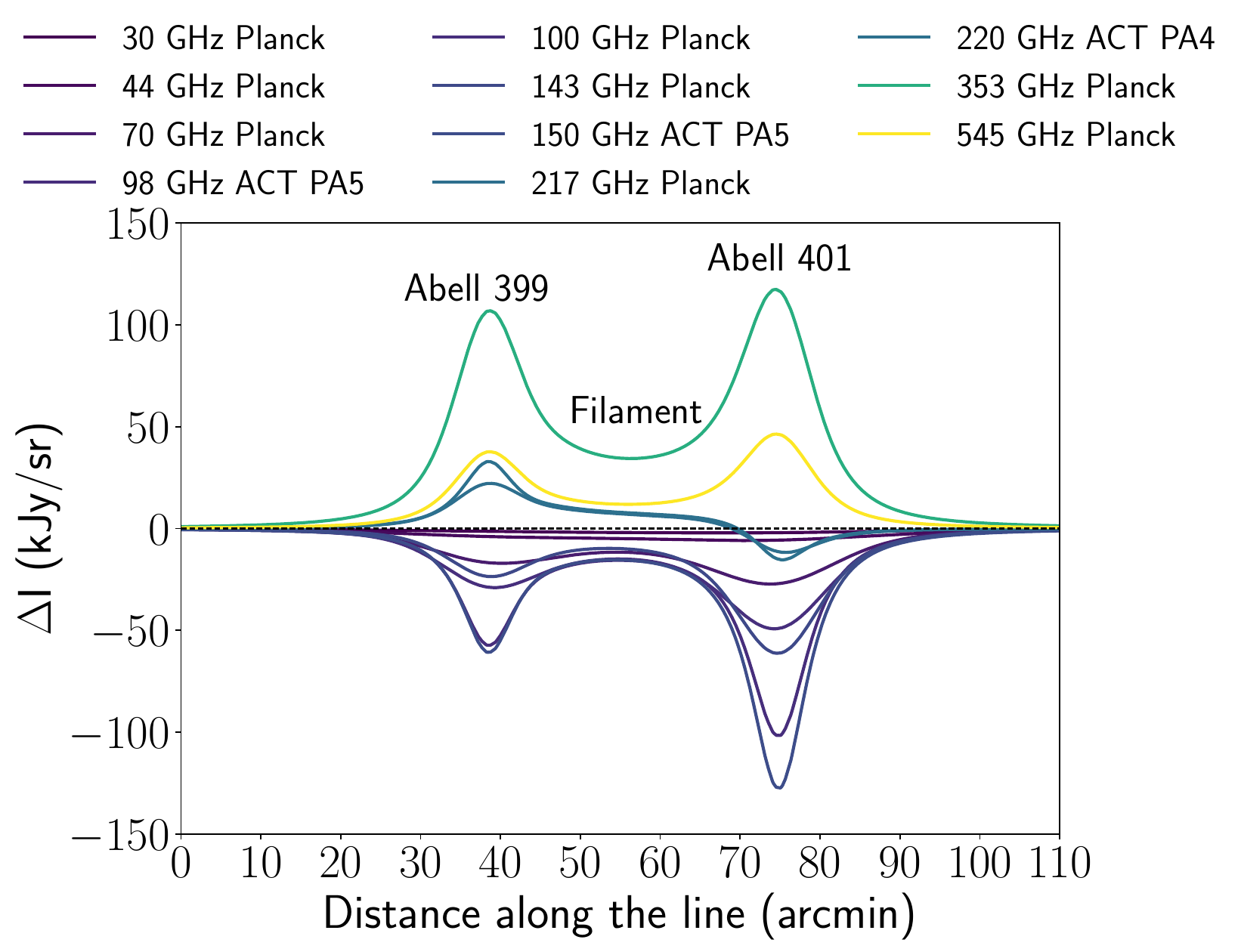}
    \caption{The best-fit SZ signal as a function of frequency along the line joining A399 and A401 for Case 1.}
    \label{fig:bridge_profile}
\end{figure}

Figures~\ref{fig:spectra}--\ref{fig:bridge_30_545} provide informative visual summaries of our results: Figure~\ref{fig:spectra} shows the spectra of the best-fit models for Case~1, including the contributions from the individual signal types---tSZ, kSZ, dust---for each of A399, A401, and the filament. Figure~\ref{fig:bridge_profile} shows a 1D slice along the axis joining A399 and A401 of the best-fit SZ model for Case~1 for a representative set of maps that were fit. Note that at frequencies near the tSZ null (\textit{Planck} 217\,GHz and ACT 220\,GHz), the kSZ signal can be seen as the curve crosses from positive to negative between the two clusters, which are moving in opposite directions. Finally, Figure~\ref{fig:bridge_30_545} provides the most complete visual representation of the results, showing the original map, the best-fit SZ model, and the residuals for each of the maps included in the fit.

The optical depth of the filament in the A399--A401 system is measured with a statistical significance of 8.5$\sigma$, providing strong evidence for its existence, in line with previous work that fit directly to Compton-$y$ maps \citep{2022Hincks, 2013PlanckFilaments, 2018Bonjean}, but now using our more flexible multifrequency method. 

The dust amplitude is not well constrained for either the clusters or the filament, in all cases being consistent with zero. The large uncertainties in the dust spectrum are not unexpected given our lack of resolution and sensitivity at higher frequencies; including better infrared data in future studies could better constrain the dust signal.

\begin{figure*}[htb!]
    \centering
    \includegraphics[width=\textwidth]{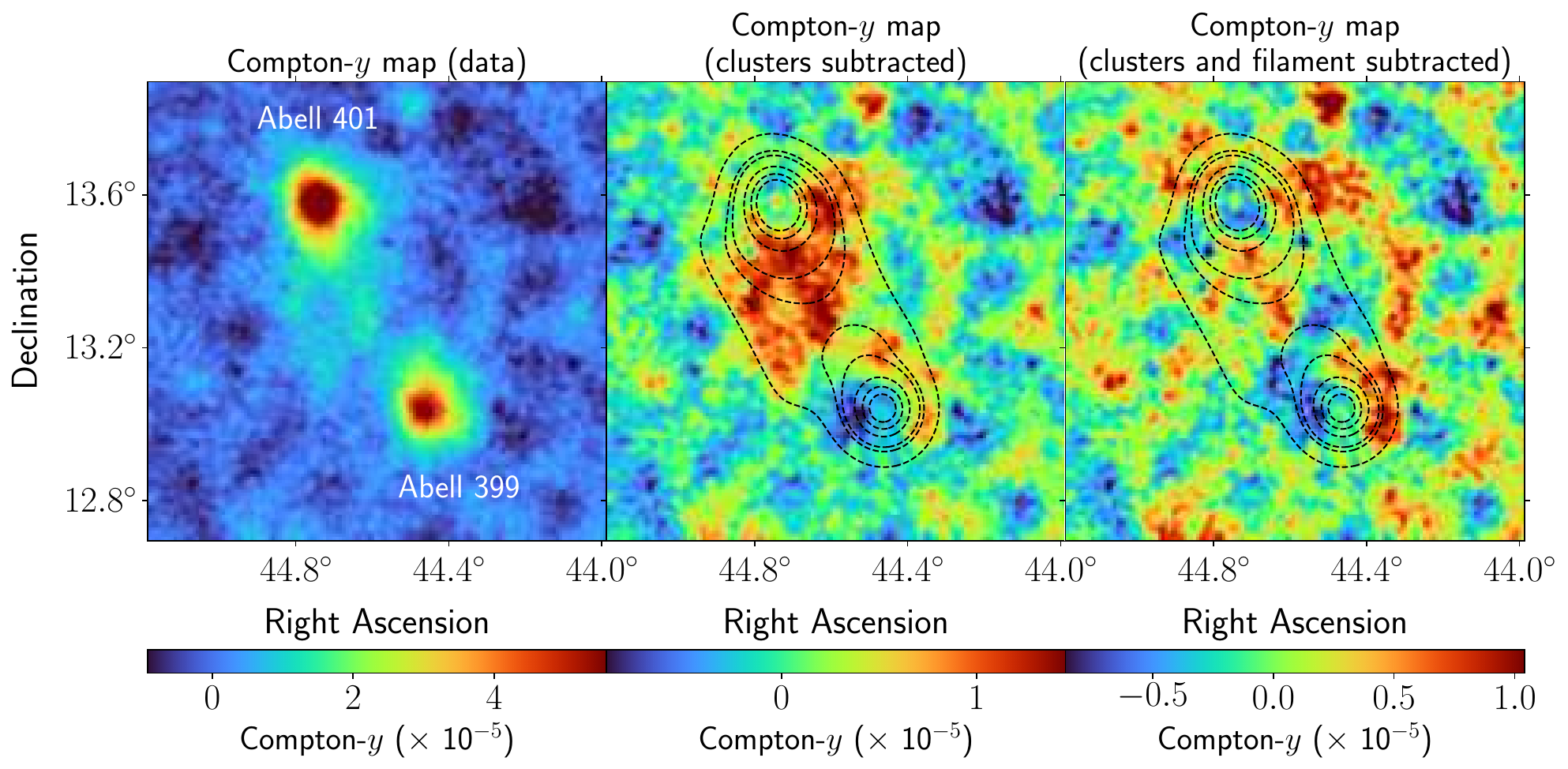}
    \caption{Compton-$y$ maps made from the multifrequency maps (left), from the residual maps with the clusters subtracted (middle), and from the residual maps with the clusters and the filament subtracted (right). The contour levels in the middle and right panel are 5$\sigma$, 4$\sigma$, 3$\sigma$, 2$\sigma$, 1.7$\sigma$, and 1$\sigma$, respectively, from the best-fit tSZ model from the PA5 98\,GHz map. The maps have been smoothed with a Gaussian filter with a standard deviation of 1.25 pixels for visualization.}
    \label{fig:compton_y_maps}
\end{figure*}

Our results exhibit the expected degeneracy between the temperature and the optical depth, since the tSZ effect is proportional to the product $n_{\rm e} T_{\rm e}$ and is only sensitive to them independently via the relativistic effect (see Equations~\ref{eqn:sz_effect} and \ref{eqn:rsz_effect}). We therefore rely on the temperature prior from X-ray measurements to break the degeneracy and obtain a good measurement of the optical depth; note that the posterior of $T_\mathrm{e}$ is virtually the same as the prior. We assess the consequences of relying on the X-ray temperature in our fits below, in Sec.~\ref{ssec:temperature_variations}. We also see a degeneracy between the optical depth and the peculiar velocity, since the kSZ effect is proportional to the product $\tau_{\rm e} v_{\rm r}$. However, because we can constrain the optical depth with good significance in the tSZ signal with the help of the X-ray temperature priors, we can obtain meaningful constraints on the peculiar velocities.

To see if any significant residuals remain after subtracting our best-fit models from the maps, we combined the residual maps shown in Figure~\ref{fig:bridge_30_545} (right columns) into a single Compton-$y$ map using the Internal Linear Combination (ILC) method \citep{2003Bennett,2009Delabrouille} as implemented in \citet{2024Coulton}. Rather than using a needlet basis, which is optimal for large maps, we work in regular Fourier space on our relatively small maps. We create two Compton-$y$ maps of residuals, shown in Figure~\ref{fig:compton_y_maps}: one in which only the best-fit model of the clusters is subtracted (leaving the bridge signal untouched) and one in which the full model, including the bridge model, is subtracted.  For visualization, we also create a Compton-$y$ of the system without any models subtracted (left panel in  Figure~\ref{fig:compton_y_maps}). The ``mesa" model appears to be a reasonable approximation of the filament, as seen in the middle panel. No obvious residuals appear in the model-subtracted maps, but there are some suggestive small-scale features in the data that are discussed in Sec.~\ref{sec:discussion}.

\subsection{Comparison to Direct Compton-$y$ Fits}
Table~\ref{tab:compare_table_y} includes the best-fit parameters from H22, who fit their model directly to a Compton-$y$ map made from \textit{Planck} data and earlier, non-deep ACT data. Comparing their results to ours is a robust check that our multifrequency method is working properly. The only significant ($>$3$\sigma$) discrepancies compared to H22 are in the right ascension, the core radius, and the $\beta$ parameter for A401. The core radius and $\beta$ are known to be correlated quantities in the isothermal $\beta$-model, so, coupled with the mismatch in right ascension, these differences from H22 are not surprising. The discrepancy in the right ascension appears to be a real difference in the maps: the H22 best-fit location for A401 from the non-dust-deprojected $y$ map appears to be off-center (see Figure \ref{fig:compton_y_map_comparison} in the Appendix). However, the best-fit coordinates of A401 from the dust-deprojected map in H22 agree with our results. This suggests that there might be a systematic astrometry mismatch in the H22 non-dust-deprojected map of around $0.5^{\prime}$ (or one pixel) in right ascension and declination. The rest of the results from our multifrequency analysis are consistent with the Compton-$y$ map results in H22. 
 
The Compton-$y$ parameter derived from our best-fit values of $n_{\rm e}$ and $T_{\rm e}$ is consistent with results obtained directly from the Compton-$y$ map in H22. This highlights the robustness of our technique in measuring the tSZ effect without relying on a $y$-map.

\subsection{Peculiar Velocities} \label{ssec:peculiar_velocity}

\begin{figure}
    \centering
    \includegraphics[width=\columnwidth]{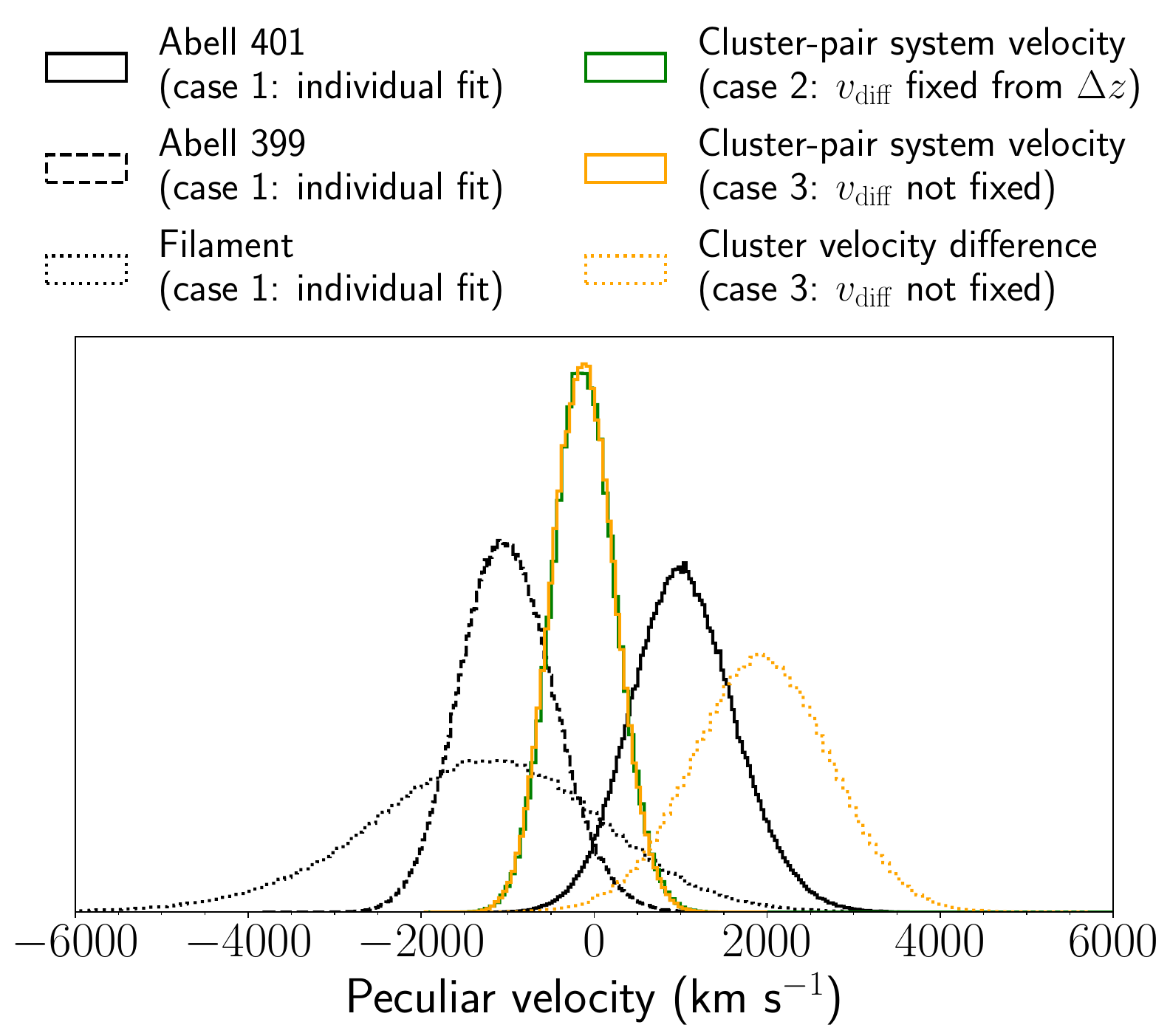}
    \caption{Comparison of peculiar velocity measurements of A399--401. The black curve shows the case where we fit the individual velocities of the two clusters and the filament (Case~1). The green curve shows the case where we fit for a single system velocity, while fixing the velocity difference between the clusters based on their redshift difference (Case~2). The orange curve shows the case where we fit for both the system velocity and the velocity difference between the two clusters (Case~3).}
    \label{fig:v_compare_fit_types}
\end{figure}

Figure \ref{fig:v_compare_fit_types} shows the posterior distributions of the velocities for the three different kSZ cases. The best-fit individual velocities in Case~1 for A401, A399, and the filament are $+1026_{-580}^{+609}\,\si{\kilo\meter\per\second}$, $-1007_{-525}^{+553}\,\si{\kilo\meter\per\second}$, and $-1252_{-1361}^{+1290}\,\si{\kilo\meter\per\second}$, respectively. The best-fit single system velocity in Case~2 is $-116_{-358}^{+376}\,\si{\kilo\meter\per\second}$, and the best-fit system velocity and the velocity difference in Case~3 are $-124_{-358}^{+366}\,\si{\kilo\meter\per\second}$ and $+1944_{-783}^{+786}\,\si{\kilo\meter\per\second}$, respectively. We measure a velocity difference of $+2033 \pm 802\,\si{\kilo\meter\per\second}$  between the two clusters in Case~1, which is consistent with the expected velocity difference ($520\,\si{\kilo\meter\per\second}$) based on the redshift difference at the 1.8$\sigma$ level. The velocity difference constrained in Case~3 is also consistent with $520\,\si{\kilo\meter\per\second}$ at the 1.8$\sigma$ level. The sign of the individual peculiar velocities indicates that A399 is moving toward us and A401 is moving away from us.

The statistical uncertainties for the individual velocities are higher than constraining a single system velocity, as expected, but are nonetheless on par with current state-of-the-art measurements of the kSZ effect in galaxy clusters (e.g., in \citealt{sayers2019}). The statistical uncertainty for the filament's individual velocity is higher than that for the individual cluster velocities, which is expected due to the lower signal-to-noise ratio of the filament signal compared to the clusters and the higher variance from the primary CMB anisotropies at the larger spatial scale of the filament compared to the individual clusters. For Cases~2 and 3, we find a consistent single system velocity and thereby appear to place a strong constraint on $v_{\mathrm{system}}$, even if we are not yet sensitive to $v_{\mathrm{diff}}$. Our results highlight the promise of this multifrequency approach to statistically constrain the kSZ effect in galaxy clusters and filaments. 

\subsection{Impact of Temperature Variations}
\label{ssec:temperature_variations}

As discussed in Sec.~\ref{ssec:posterior_distributions}, we are not able to independently constrain the gas density and gas temperature from SZ measurement alone, and rely on X-ray-derived temperatures to break this degeneracy. This potentially introduces a bias in our fits because the X-ray brightness arises from free--free emission and is proportional to $n_{\mathrm{e}}^2$, whereas the SZ effects are proportional to $n_{\mathrm{e}}$. Thus, the temperature obtained from X-ray spectroscopy has a different weighting of $n_{\mathrm{e}}$ along the line of sight than the weighting relevant to the SZ signal \citep{vikh2009, 2020ELEE, 2017adam_xray, 2024kay}. 

Using FLAMINGO simulations, \citet{2024kay} found that the difference between the $y$-weighted temperature (relevant to the SZ) and the X-ray spectroscopic temperature is $\sim$5\% from $r = 0.1R_{500c}$ to $r = 0.5R_{500c}$, with the difference increasing further starting at $r = 0.5R_{500c}$ to up to 25\% at $r = R_{500c}$, where $R_{500c}$ is the radius within which the average density is 500 times the critical density of the Universe at the cluster's redshift. To investigate whether this difference in temperature measures significantly biases our result, we varied the central value of the Gaussian prior of the cluster temperature from 5\% to 25\% away from the X-ray-based values, and observed how this altered the best-fit peculiar velocity, optical depth, and the Compton-$y$ parameter.

\begin{figure}
    \centering
    \begin{minipage}{0.49\textwidth}
        \centering
        \includegraphics[width=\linewidth]{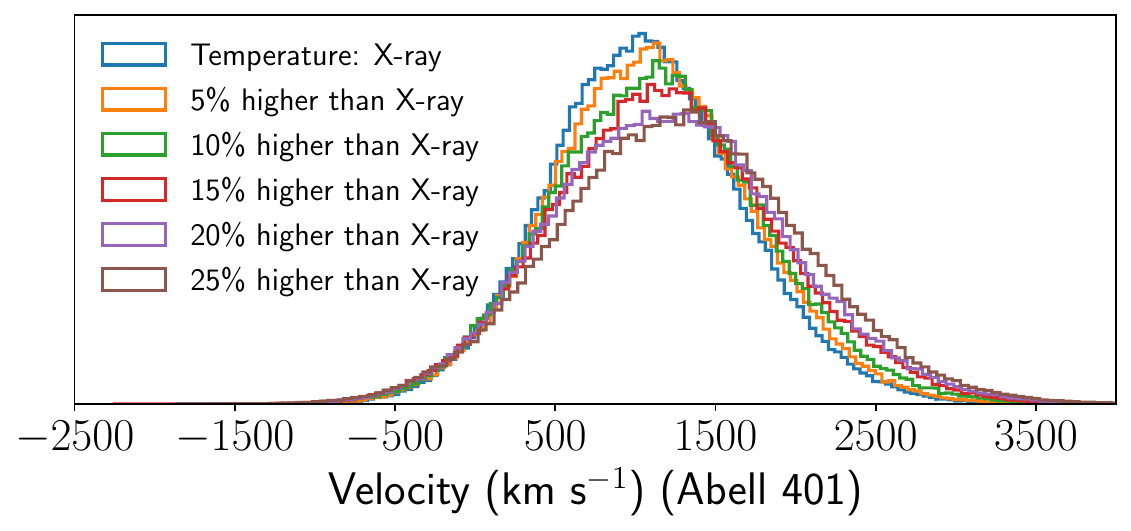}
    \end{minipage}
    \hfill
    \begin{minipage}{0.49\textwidth}
        \centering
        \includegraphics[width=\linewidth]{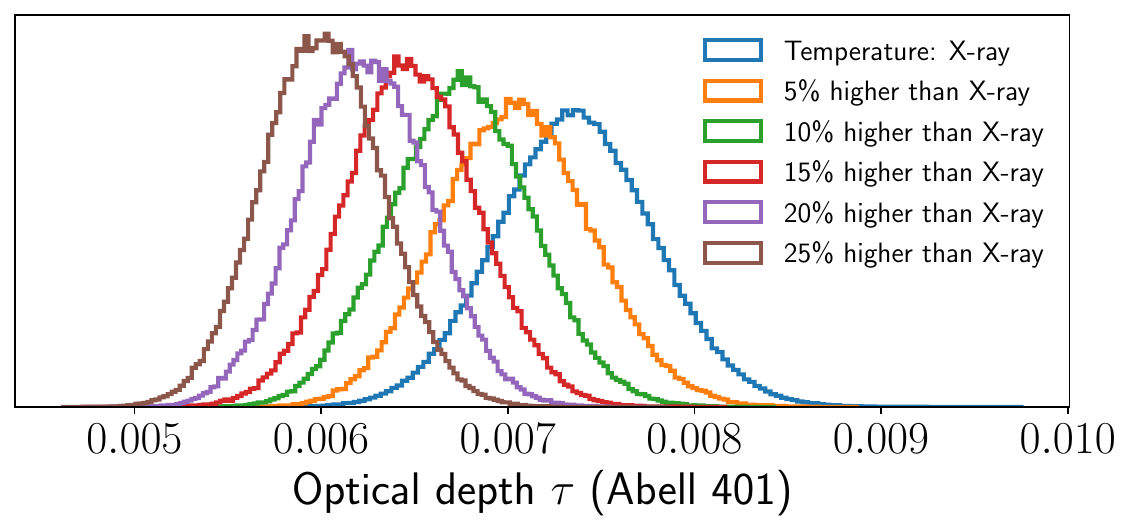}
    \end{minipage}
    \hfill
    \begin{minipage}{0.49\textwidth}
        \centering
        \includegraphics[width=\linewidth]{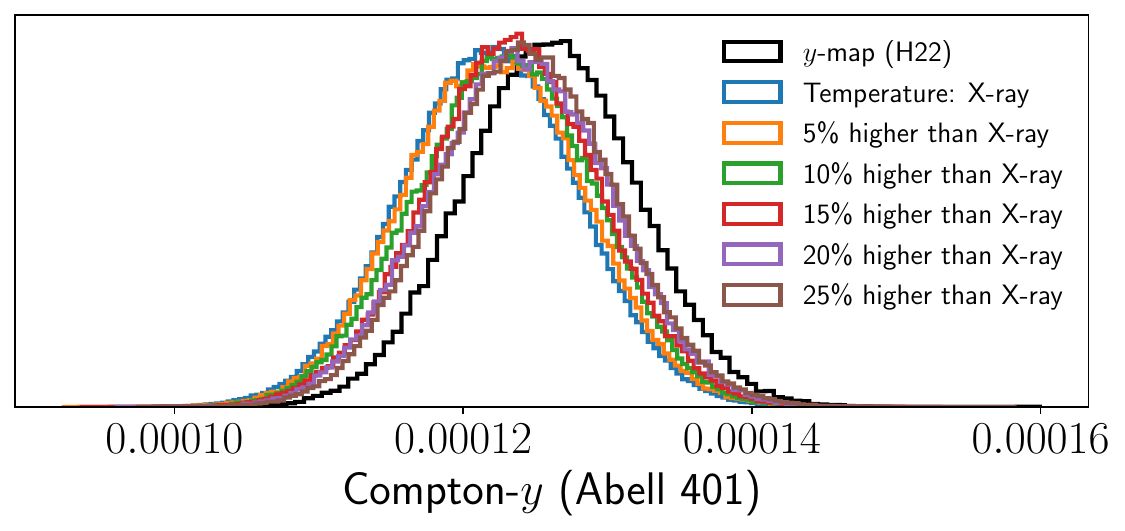}
        \label{fig:y_Te_ind_vc_a401}
    \end{minipage}
    \caption{Constraints on A401 best-fit parameters as a function of gas temperature. The distribution in black in the Compton-$y$ map (bottom) is from \citet{2022Hincks}}
            \label{fig:Te_ind_vc_a401}
\end{figure}

The top panel in Figure~\ref{fig:Te_ind_vc_a401} shows the peculiar velocity of A401 as a function of temperature. Although the error bars remain large, there is a shift in the best-fit value of the velocity in the direction expected for preserving the tSZ and kSZ signals. As the temperature increases, the best-fit optical depth decreases, as shown in the middle panel of Figure~\ref{fig:Te_ind_vc_a401}. These results are not surprising, since the optical depth and the temperature are degenerate in the tSZ signal. As the optical depth decreases, the magnitude of the best-fit peculiar velocity increases to preserve the kSZ signal. However, even up to temperature variations of 25\%, we find that the statistical uncertainty in the peculiar velocity dominates compared to temperature systematics. 
We also investigated the difference in the Compton-$y$ parameter with temperature variations. The bottom panel in Figure~\ref{fig:Te_ind_vc_a401} shows the Compton-$y$ parameter distribution as a function of temperature for A401 under Case~1 (fitting its individual velocity). There is no significant change in the best-fit Compton-$y$ parameter with temperature, indicating the tSZ signal is effectively being separated from the kSZ signal with this methodology. Comparing with the $y$-map results from H22, we find a $0.42\sigma$, $1.21\sigma$, and $1.32\sigma$ discrepancy in the $y$ parameter for A401, A399, and the filament, respectively. While the discrepancy in Compton-$y$ for the filament is larger than the clusters, it remains statistically insignificant. The result of temperature variations on the kSZ effect and the Compton-$y$ parameter were similar for A399 and the filament and for the other two kSZ cases, so we only show the results for A401 under Case~1 for brevity.

\section{Discussion}\label{sec:discussion}
  
\begin{figure*}
    \centering
    \includegraphics[width=\textwidth]{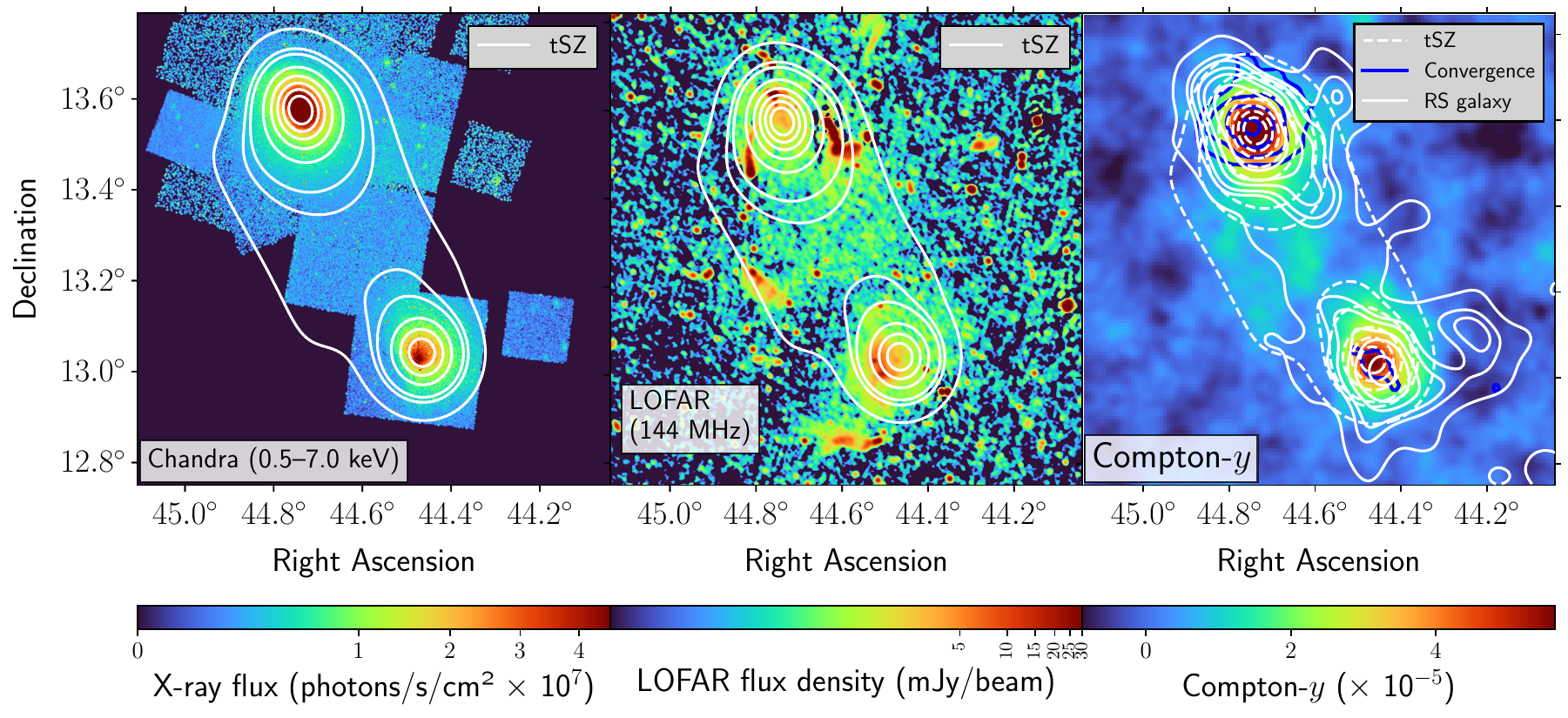}
    \caption{\textit{Chandra} X-ray map (left), LOFAR map (middle), and Compton-$y$ map (right) of the A399--A401 system. We show best-fit tSZ contours from our PA5 98 GHz model at 1$\sigma$, 1.7$\sigma$, 2$\sigma$, 3$\sigma$, 4$\sigma$, 5$\sigma$, 6$\sigma$, and 7$\sigma$ levels.
    We also show the convergence and red sequence (RS) galaxy distribution contours taken from \citet{2024Fu} over the $y$ map. The contour levels for the convergence map are 2$\sigma$, 3$\sigma$, 4$\sigma$, and 5$\sigma$. The levels for the RS distribution are 1$\sigma$, 1.7$\sigma$, 2$\sigma$, 2.5$\sigma$, 3$\sigma$, and 4$\sigma$.} 
    \label{fig:multi_wave_figure_sep2}
\end{figure*}

The left panel of Figure \ref{fig:multi_wave_figure_sep2} shows our best-fit tSZ contours from the ACT PA5 98 GHz model overlaid on the \textit{Chandra} X-ray map. The X-ray emission shows broad overall agreement with the best-fit model for the tSZ effect in the system. We note, however, that A399 exhibits an offset between the tSZ and X-ray signals typical of a merger \citep{2014Zhang}.  The merger hypothesis in A399 is well-established and supported by multiple previous studies \citep{2004Sakelliou, 2008Bourdin, 2010Murgia}, while closer examination shows a large-scale sloshing spiral likely induced by the same merger producing the offset.  We also note that the connection of the bridge to A399, rather than following the direct linear path between the cluster's X-ray peaks, appears to be offset farther to the west, consistent with the direction of the tSZ/X-ray offset within A399. This feature in the bridge is even more apparent in the Compton-$y$ residual map showing the bridge signal only (middle panel of Figure~\ref{fig:compton_y_maps}). The residuals perhaps show hints of smaller-scale morphology in the bridge that are not captured by the simple mesa model, such as the middle of the bridge, on the east side, where the signal is strongest, consisting of two ``fingers'', but the signal-to-noise of such features is too low to say anything conclusive.  

In the middle panel of Figure \ref{fig:multi_wave_figure_sep2}, the best-fit tSZ model is shown on the LOFAR radio imaging. We note the strong correspondence between the diffuse radio emission and the thermal gas \citep[as already seen in, e.g.,][]{2019Govoni, 2022Radiconi}.  
The arc-like radio feature to the southeast of A399, visible in the LOFAR observations, was identified as a possible radio relic by \cite{2019Govoni, 2022Radiconi} that may be associated with the merger component driving the sloshing event. However, examination of high-resolution LOFAR observations \citep{deJong2022} indicates that it is more likely a radio galaxy that may simply be located near a sloshing discontinuity in projection.
Overall, further work will be required to provide a more detailed SZ+X-ray characterization of the merger and sloshing activity in A399. 

The right-hand panel of Figure \ref{fig:multi_wave_figure_sep2} shows the Compton-$y$ map with lensing convergence and red sequence (RS) galaxy density contours overlaid from \citet{2024Fu} using data from the Dark Energy Camera. While the peaks in the convergence map appear to agree well, within the uncertainties, with the corresponding peaks of the tSZ map, there is an apparent offset between the RS galaxy distribution peak and the tSZ in A399 that may further hint that this system is out of thermal hydrostatic equilibrium. Further, we highlight the excess overdensity of RS galaxies in the filament region, as previously reported in \citealt{2024Fu}. There are also hints of correlation between the RS galaxy density and the tSZ signal at the link regions connecting the filament with the clusters, confirming that the offset is likely real. Again, further study of the exact nature of the correspondence and morphological offset is merited.

\section{Conclusion}
In this work, we present a multifrequency and multi-instrument approach to study the physical properties of galaxy clusters and filaments using high-resolution cosmic microwave background data. This technique allows for measuring the thermal and kinematic Sunyaev-Zeldovich effects in clusters and filaments, while accounting for relativistic corrections. In contrast to the previous technique of using Compton-$y$ maps to measure the tSZ effect, this approach provides more flexibility in accounting for both relativistic corrections to the tSZ and kSZ effect, and allows modelling the astrophysical foreground, such as fitting the dust emission and point sources as needed in individual systems on a per-pixel basis. We present the methodology for this approach and apply it to the A399--401 cluster pair system using maps from the Atacama Cosmology Telescope and the \textit{Planck} satellite. We assumed Gaussian priors on the gas temperature based on X-ray data. 

We measure the filament optical depth with $\sim$$8.5\sigma$ precision and find tSZ results consistent with previous studies of the system using Compton-$y$ maps. We also measure the individual peculiar velocities of the two clusters and the filament, obtaining statistical uncertainties of $\lesssim\,$600\,km\,s$^{-1}$, competitive with current state-of-the-art measurements. We also explore the impact of temperature variation on the recovered parameters.

Our approach is well-suited to apply to future instruments across subterahertz frequencies to study galaxy clusters and filaments, such as, for example, the Simons Observatory  \citep[SO;][]{2025SOCollab}, the CCAT Observatory's Fred Young Submillimeter Telescope \citep[FYST;][]{CCAT-p2023}, CMB-HD \citep[CMB-HD;][]{CMB-HD-2022}, and the Atacama Large Aperture Submillimeter Telescope \citep[AtLAST;][]{2025DiMascolo}.

\begin{acknowledgments}
We thank Craig Sarazin, Luca Di Mascolo, Jens Chluba, Elizabeth Lee, Jonathan Sievers, and Kerri Cahoy for helpful discussions. We thank Shenming Fu for providing us with the convergence and red sequence galaxy distribution maps for A399--401.

Support for ACT was through the U.S.~National Science Foundation through awards AST-0408698, AST-0965625, and AST-1440226 for the ACT project, as well as awards PHY-0355328, PHY-0855887, and PHY-1214379. Funding was also provided by Princeton University, the University of Pennsylvania, and a Canada Foundation for Innovation (CFI) award to UBC. ACT operated in the Parque Astron\'omico Atacama in northern Chile under the auspices of the Agencia Nacional de Investigaci\'on y Desarrollo (ANID). The development of multichroic detectors and lenses was supported by NASA grants NNX13AE56G and NNX14AB58G. Detector research at NIST was supported by the NIST Innovations in Measurement Science program. 
\end{acknowledgments}

Computing for ACT was performed using the Princeton Research Computing resources at Princeton University and the Niagara supercomputer at the SciNet HPC Consortium. SciNet is funded by the CFI under the auspices of Compute Canada, the Government of Ontario, the Ontario Research Fund–Research Excellence, and the University of Toronto \citep{scinet1, scinet2}. This research also used resources of the National Energy Research Scientific Computing Center (NERSC), a U.S. Department of Energy Office of Science User Facility located at Lawrence Berkeley National Laboratory, operated under Contract No. DE-AC02-05CH11231 using NERSC award HEP-ERCAPmp107 from 2021 to 2025. 

\begin{acknowledgments}
    We thank the Republic of Chile for hosting ACT in the northern Atacama, and the local indigenous Licanantay communities, whom we follow in observing and learning from the night sky. YG acknowledges support from the University of Toronto's Eric and Wendy Schmidt AI in Science Postdoctoral Fellowship, a program of Schmidt Sciences. The Dunlap Institute is funded through an endowment established by the David Dunlap family and the University of Toronto. ADH acknowledges support from the Sutton Family Chair in Science, Christianity and Cultures, from the Faculty of Arts and Science, University of Toronto, and from the Natural Sciences and Engineering Research Council of Canada (NSERC) [RGPIN-2023-05014, DGECR-2023-00180].  AG acknowledges support from the NASA CLICK program award (80NSSC18K1579), NASA Astrophysics Research and Analysis (APRA) award (80NSSC24K1558), and NASA Strategic Astrophysics Technology (SAT) award (80NSSC23K0350). TM acknowledges support from the Agencia Estatal de Investigaci\'on (AEI) and the Ministerio de Ciencia, Innovaci\'on y Universidades (MICIU) Grant ATRAE2024-154740 funded by MICIU/AEI//10.13039/501100011033. KM acknowledges support from the National Research Foundation of South Africa. CS acknowledges support from the Agencia Nacional de Investigaci\'on y Desarrollo (ANID) through Basal project FB210003.

\software{\texttt{Astropy} \citep{astropy:2013, astropy:2018, astropy:2022},
            \texttt{SZPack} \citep{CNSN, CSNN},
            \texttt{Pixell}\citep{2021pixell},
          \texttt{SAOImage DS9} \citep{ds9},
          \texttt{emcee} \citep{emcee},
          \texttt{NumPy} \citep{harris2020array},
          \texttt{SciPy} \citep{2020SciPy-NMeth},
                    \texttt{numba} \citep{lam2015numba},
          \texttt{Matplotlib} \citep{Hunter:2007}.
          }
\end{acknowledgments}

\begin{figure*}[htb!]
    \centering
    \includegraphics[width=\textwidth]{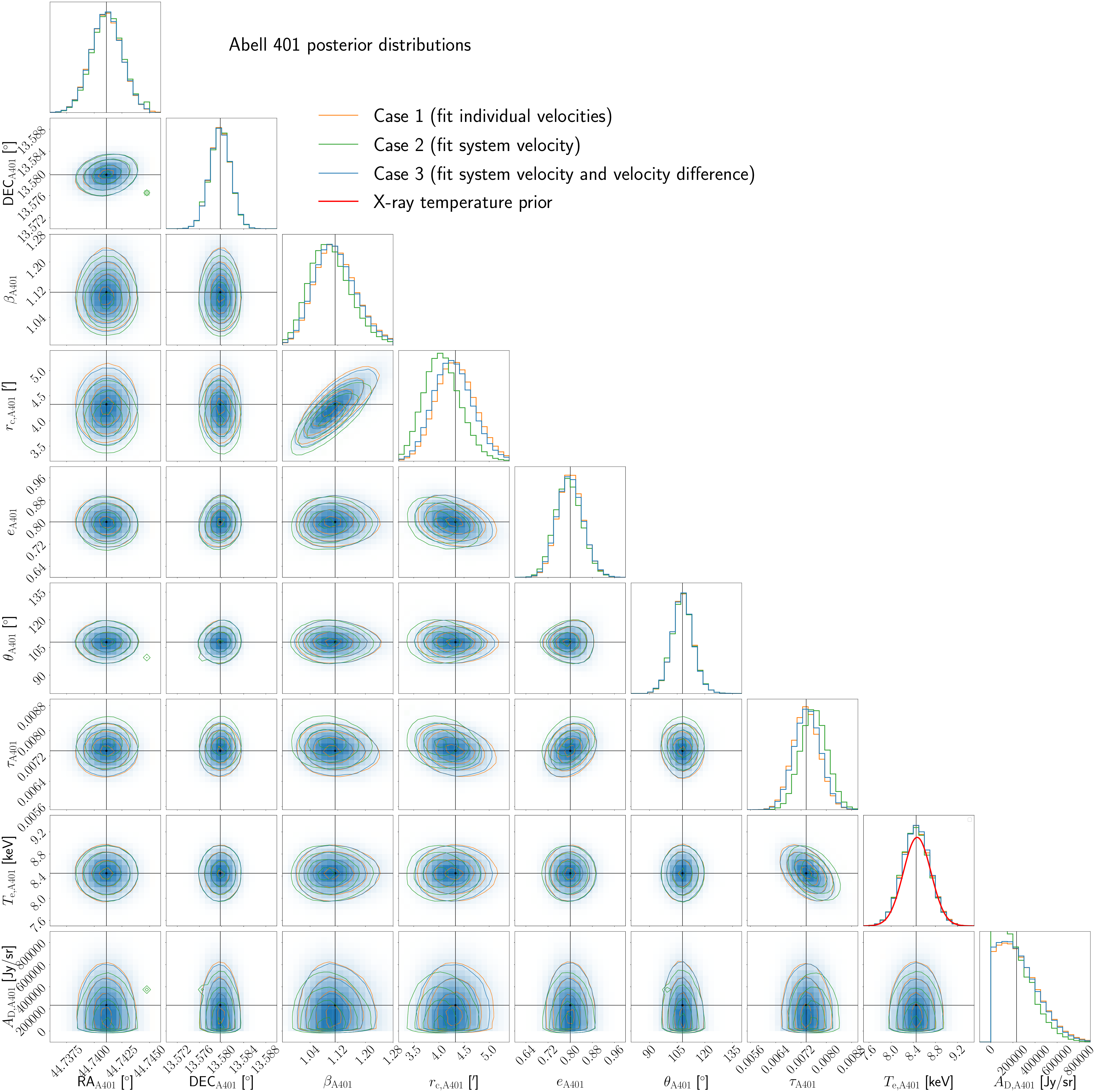}
    \caption{Corner plots for cases 1, 2, and 3 for Abell 401. The contour levels are 0.5$\sigma$, 1$\sigma$, 1.5$\sigma$, and 2$\sigma$. The red curve shows the Gaussian prior used for the gas temperature.}
    \label{fig:401_corner}
\end{figure*}

\begin{figure*}[htb!]
    \centering
    \includegraphics[width=\textwidth]{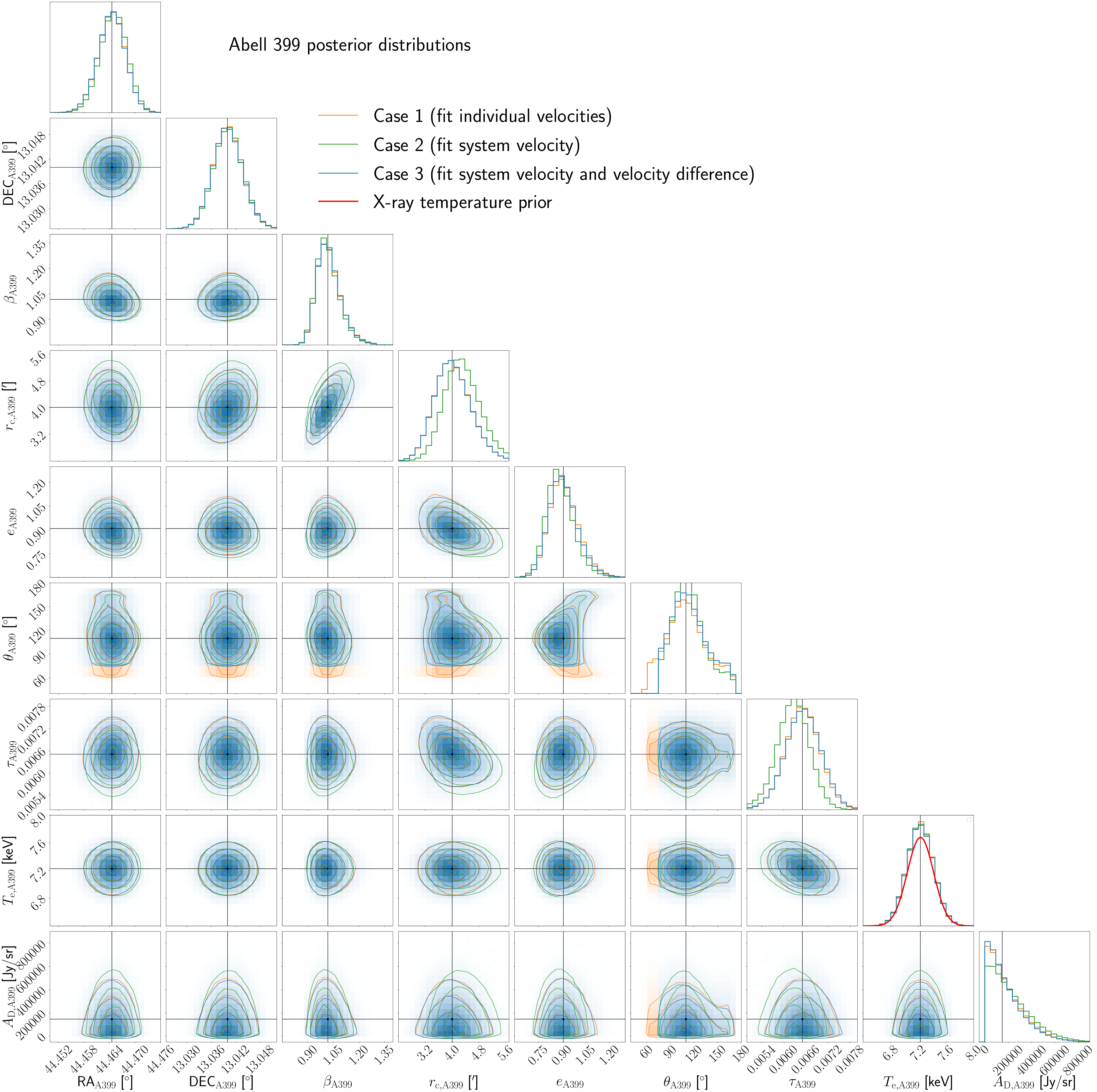}
    \caption{Corner plots for cases 1, 2, and 3 for Abell 399. The contour levels are 0.5$\sigma$, 1$\sigma$, 1.5$\sigma$, and 2$\sigma$. The red curve shows the Gaussian prior used for the gas temperature.}
    \label{fig:399_corner}
\end{figure*}

\begin{figure*}[htb!]
    \centering
    \includegraphics[width=\textwidth]{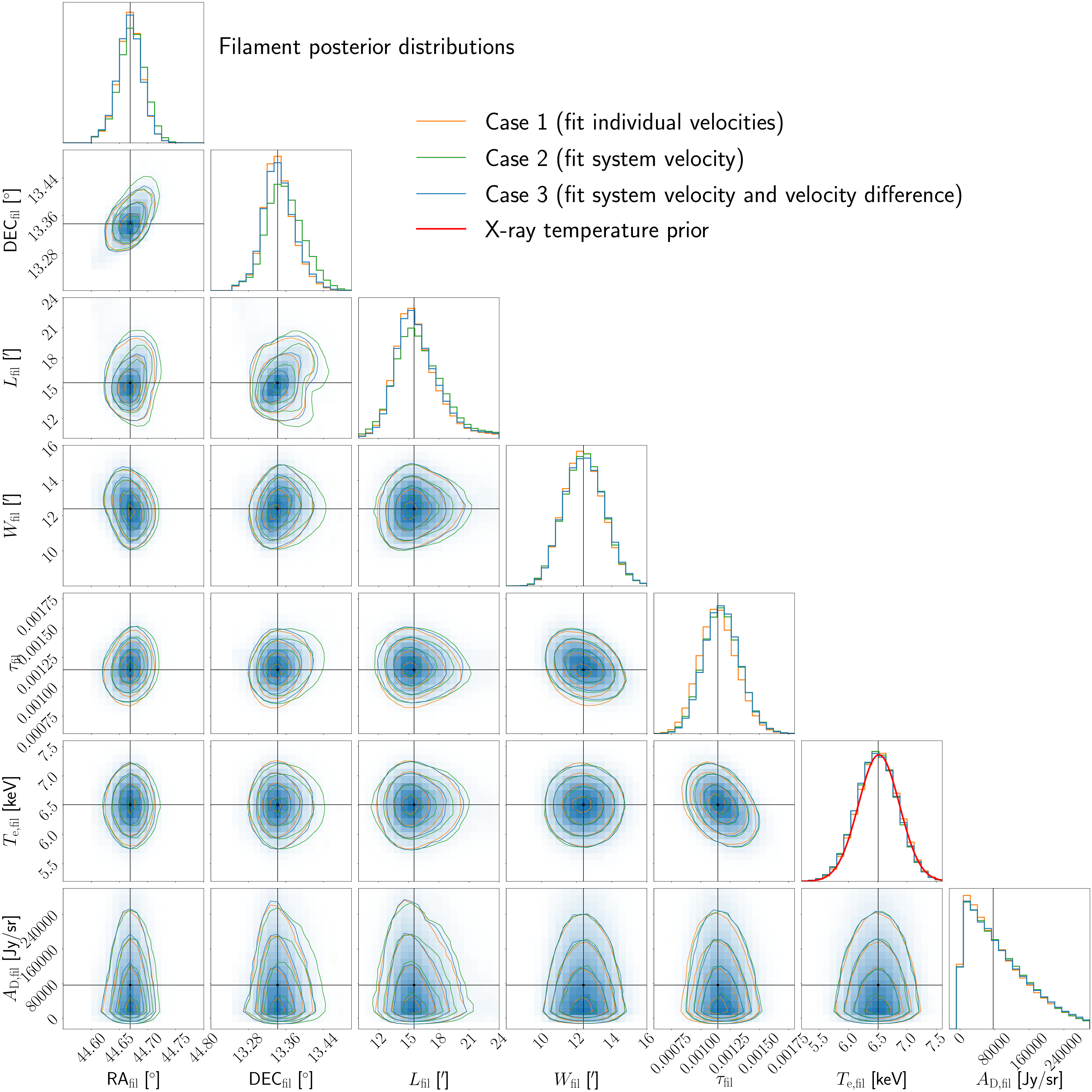}
    \caption{Corner plots for cases 1, 2, and 3 for the filament. The contour levels are 0.5$\sigma$, 1$\sigma$, 1.5$\sigma$, and 2$\sigma$. The red curve shows the Gaussian prior used for the gas temperature.}
    \label{fig:fil_corner}
\end{figure*}

\begin{figure}[htb!]
    \centering
    \includegraphics[width=\columnwidth]{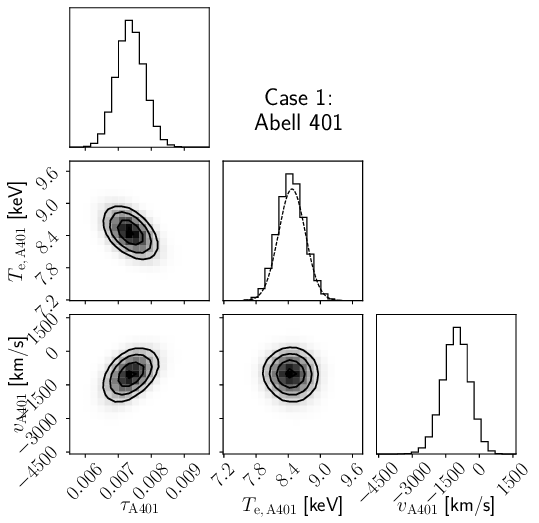}
    \caption{The posterior distribution of the optical depth, the gas temperature, and the peculiar velocity for Abell 401.}
    \label{fig:abell401_corner_tau_v_T}
\end{figure}

\begin{figure}[htb!]
    \centering
    \includegraphics[width=\columnwidth]{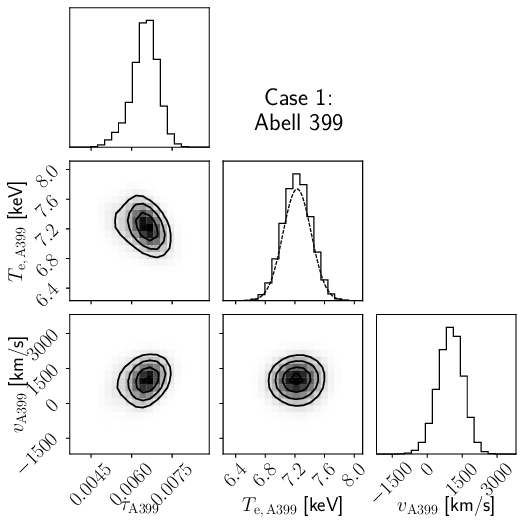}
    \caption{The posterior distribution of the optical depth, the gas temperature, and the peculiar velocity for Abell 399.}
    \label{fig:abell399_corner_tau_v_T}
\end{figure}

\begin{figure}[htb!]
    \centering
    \includegraphics[width=\columnwidth]{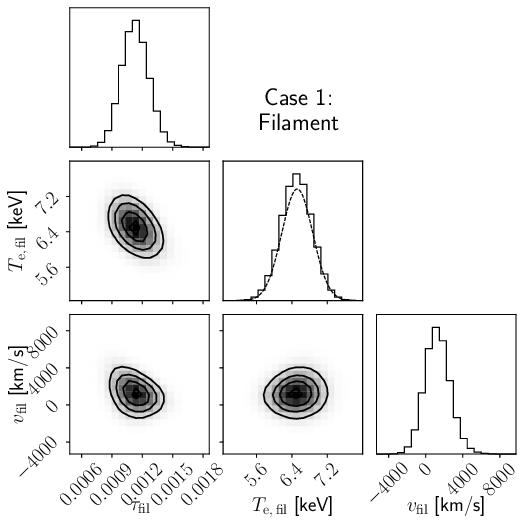}
    \caption{The posterior distribution of the optical depth, the gas temperature, and the peculiar velocity for the filament.}
    \label{fig:filament_corner_tau_v_T}
\end{figure}

\begin{figure}[htb!]
    \centering
    \begin{minipage}{0.48\textwidth}
        \centering
        \includegraphics[width=0.9\linewidth]{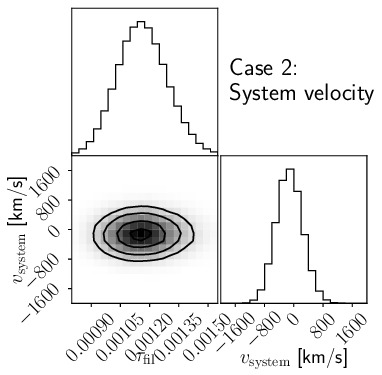}
    \end{minipage}
    \hfill
    \begin{minipage}{0.48\textwidth}
        \centering
        \includegraphics[width=0.6\linewidth]{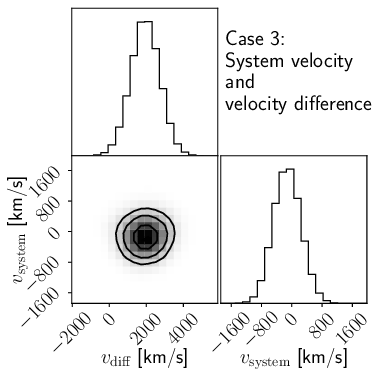}
    \end{minipage}
        \caption{The system velocity and the velocity difference posterior distributions for Case~2 and Case~3.}
    \label{fig:v_cases2_and_3}
\end{figure}

\begin{figure}[htb!]
    \centering
    \includegraphics[width=.8\columnwidth]{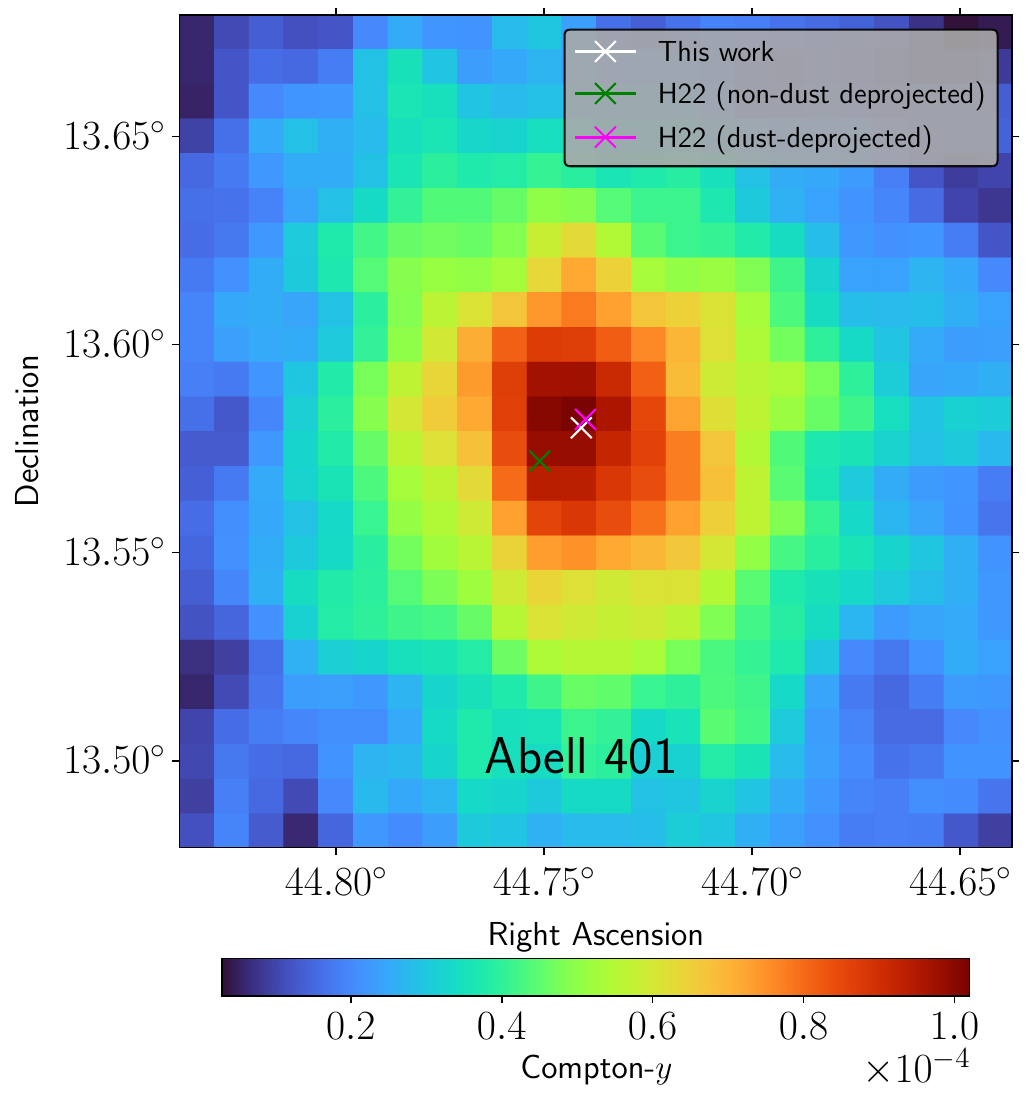}
    \caption{This shows the Compton-$y$ of Abell 401 made from the same maps we used in this work. The discrepancy between the best-fit location of Abell 401 using this multifrequency approach and the Compton-$y$ maps in \citealt{2022Hincks} appears to be due to a systematic offset in the astrometry ($\sim 0.5^{\prime}$ or one pixel in Right Ascension and Declination) in the non-dust-deprojected Compton-$y$ map.}
    \label{fig:compton_y_map_comparison}
\end{figure}

\bibliography{main}
\appendix
In this appendix, we include the full corner plots for our fit results. There are too many parameters to show in a single plot, so we split them up as follows: all parameters except the peculiar velocity are shown for A401 in Figure~\ref{fig:401_corner}, for A399 in Figure~\ref{fig:399_corner}, and for the filament in Figure~\ref{fig:fil_corner}. The optical depth, temperature, and velocity---the three parameters that determine the amplitude of the tSZ and kSZ signals---are shown for A401 in Figure~\ref{fig:abell401_corner_tau_v_T}, for A399 in Figure~\ref{fig:abell399_corner_tau_v_T}, and for the filament in Figure~\ref{fig:filament_corner_tau_v_T}, all for Case~1. The velocities for Case~2 and Case~3 are shown in Figure~\ref{fig:v_cases2_and_3}. Finally, we show the discrepancy between our results and H22 for the best-fit position of A401 in Figure~\ref{fig:compton_y_map_comparison}.

\end{document}